\documentclass[epj,nopacs,linenumbers]{svjour}
\usepackage{cuted}
\usepackage{graphics}
\usepackage{graphicx}  
\usepackage{dcolumn}   
\usepackage{bm}        
\usepackage{amssymb}   
\usepackage{dsfont}
\usepackage{xcolor}
\usepackage{nccfoots}
\usepackage{epsfig}
\usepackage{epstopdf}

\definecolor{eq}{rgb}{190,190,190}
\usepackage{amsmath}
\usepackage[latin9]{inputenc}
\usepackage[flushleft]{threeparttable}
\usepackage{color}

\def\vA{\vec{A}}

\def\H{\hat{H}}

\def\hvj{\hat{\vec{j}}}
\def\k{\vec{k}}
\def\LL{\mathcal{L}}
\def\r{\vec{r}}
\def\hro{\hat{\rho}}
\def\sig{\hat{\sigma}}

\newcommand{\ket}[1]{\left| #1 \right\rangle}

\newcommand{\bra}[1]{\left\langle #1 \right|}

\begin{document}
\title{The theory of direct laser excitation of nuclear transitions}

\author{Lars von der Wense$^{*}$\inst{1}  \and Pavlo V. Bilous\inst{2} \and Benedict Seiferle\inst{1}  \and Simon Stellmer\inst{3} \and Johannes Weitenberg\inst{4} \and Peter G. Thirolf\inst{1}  \and Adriana Pálffy$^{**}$\inst{2} \and Georgy Kazakov\inst{5,6}}

\institute{LMU Munich, 85748 Garching, Germany. \and Max Planck Institute for Nuclear Physics, 69117 Heidelberg, Germany. \and University of Bonn, 53105 Bonn, Germany. \and Max Planck Institute of Quantum Optics, 85748 Garching, Germany. \and WPI c/o Fak. Mathematik Univ. Wien, Oskar-Morgenstern-Platz 1, 1090 Vienna, Austria. \and Atominstitut, TU Wien, Stadionallee 2, 1020 Vienna, Austria.}
%
%
\abstract{
A comprehensive theoretical study of direct laser excitation of a nuclear state based on the density matrix formalism is presented. The nuclear clock isomer $^{229\text{m}}$Th is discussed in detail, as it could allow for direct laser excitation using existing technology and provides the motivation for this work. The optical Bloch equations are derived for the simplest case of a pure nuclear two-level system and for the more complex cases taking into account the presence of magnetic sub-states, hyperfine-structure and Zeeman splitting in external fields. Nuclear level splitting for free atoms and ions as well as for nuclei in a solid-state environment is discussed individually. Based on the obtained equations, nuclear population transfer in the low-saturation limit is reviewed. Further, nuclear Rabi oscillations, power broadening and nuclear two-photon excitation are considered. Finally, the theory is applied to the special cases of $^{229\text{m}}$Th and $^{235\text{m}}$U, being the nuclear excited states of lowest known excitation energies. The paper aims to be a didactic review with many calculations given explicitly.}
\maketitle

\section{\label{sec1}Introduction}
\noindent Direct excitation of a nuclear state using narrow-bandwidth laser radiation remains an outstanding experimental challenge. When achieved, potential applications in various physical fields open up. These include the development of a highly stable source of light for metrology \cite{Tkalya1}, the development of a nuclear optical clock \cite{Peik1,Campbell2}, a nuclear $\gamma$-ray laser \cite{Rivlin,Tkalya2} and a nuclear qubit for quantum computing \cite{Gunst}. Further, it will advance the field of experimental nuclear quantum optics \cite{Buervenich,Das,Liao2}. With an expected accuracy of $1.5\cdot10^{-19}$ \cite{Campbell2} a nuclear optical clock may even surpass the best optical atomic clocks operational today \cite{McGrew,Brewer}, having the potential for utilization in, e.g., satellite-based navigation \cite{Thirolf}, relativistic geodesy \cite{Mehlstaeubler}, probing for time-variations of fundamental constants \cite{Flambaum} and the search for topological dark matter \cite{Derevianko}.\Footnotetext{}{$^*$l.wense@physik.uni-muenchen.de} \Footnotetext{}{$^{**}$palffy@mpi-hd.mpg.de}\\[0.2cm]
Two different concepts of nuclear laser excitation have to be distinguished: direct excitation and excitation via the inverse electronic bridge (IEB) mechanism. In the IEB process, a virtual level of the electronic shell is excited and the energy is subsequently transferred to the nucleus. It was proposed for laser excitation of $^{229\text{m}}$Th in Refs.~\cite{Tkalya1992a,Tkalya1992b} and has since then been discussed in various publications \cite{Tkalya1,Kalman1994,Typel1996,Matinyan,Karpeshin1999,Karpeshin2002,Porsev2010,Romanenko,Karpeshin2015,Karpeshin2017,Bilous2018,Andreev2019,Mueller2019,Borisyuk2019,Bilous2020,Nickerson2020}. Generally, the IEB process is expected to result in larger nuclear excitation rates than the direct laser excitation. Here, however, purely direct nuclear laser excitation is discussed, allowing for a complete analytical treatment.\\[0.2cm]
Despite the interesting expected applications, narrowband laser excitation of a nuclear state has not yet been achieved. The central reason is that conventional laser technology is making important use of the valence electrons in the atomic shell, with energies in the range of up to a few eV, while the typical nuclear energy scale extends to significantly higher energies, ranging from a few keV to up to several MeV. Thus there is no overlap between the nuclear energy scale and conventional laser technology, posing an inherent problem for direct nuclear laser excitation. One way out could be to use different ways of laser light generation, e.g. via high-harmonic generation (HHG) \cite{Gohle,Cingoez,Pupeza,Porat,Zhang,Saule} or the use of free-electron lasers \cite{McNeil,Allaria}. Also X-ray pulse shaping that involves highly charged ions has been proposed \cite{Cavaletto}.\\[0.2cm]
In fact, X-ray free-electron lasers (XFELs) already allow to generate coherent light of up to 20~keV in energy, thereby covering some of the low-energy nuclear excited states \cite{Ueda}. This has led to the first successful XFEL-excitation of a nuclear state \cite{Chumakov} and is expected to advance the already established field of nuclear quantum optics \cite{Buervenich,Roehlsberger,Haber}. Compared to the extremely narrow bandwidths in the sub-Hz range that can be generated with atomic-shell-based laser technology, the XFEL technology with bandwidths in the range of a few eV can, however, be considered as broad-band. For this reason the technology is not suitable for the development of a nuclear clock.\\[0.2cm]
In contrast, the technology of HHG has been used to generate extreme ultraviolet (XUV) light of up to 100~eV at a high repetition rate, as required for frequency combs \cite{Gohle,Cingoez,Pupeza,Porat,Zhang}. However, the covered XUV energy range is not sufficient in order to access most of the nuclear transitions even of low energy.\\[0.2cm]
There are two confirmed exceptions of nuclear states with extraordinary low excitation energy. These are $^{229\text{m}}$Th (8.3 eV) and $^{235\text{m}}$U (76.7 eV). Here the ``m" is short for ``meta-stable", indicating that the nuclear excited state possesses a lifetime longer than a nanosecond. Of these two nuclear states, only $^{229\text{m}}$Th offers a realistic chance for direct nuclear laser excitation, as will be discussed at the end of this paper. One further potential nuclear excited state of low excitation energy, $^{229\text{m}}$Pa \cite{Ahmad}, remains elusive and will not be discussed in this paper. In the following, the history of $^{229\text{m}}$Th will briefly be sketched.\\[0.2cm]
\noindent $^{229}$Th was first considered to possess a low-energy excited state in 1976, when it was found that its existence enabled a consistent picture to be built on the spectroscopic observations \cite{Kroger}. At that time the excitation energy of this ``thorium isomer" could only be estimated to be below 100~eV. However, by more than a decade long experimental efforts, the energy was first constrained to $-1\pm4$~eV in 1990 \cite{Reich} and later measured to be $3.5\pm1$~eV in 1994 \cite{Helmer}.\\[0.2cm]
The existence of a nuclear excited state of below 10~eV excitation energy soon afterward raised some interest from theory \cite{Strizhov} and a first quantitative analysis of the probability for direct nuclear laser excitation of $^{229\text{m}}$Th was published as early as 1992 \cite{Tkalya1992b}. Performing nuclear laser spectroscopy was proposed by Peik and Tamm in 2003 together with the proposal for the development of a nuclear optical clock \cite{Peik1}. Two different types of clocks have since then been discussed: the single ion nuclear clock, which is based on an individual $^{229}$Th$^{3+}$ ion in a Paul trap \cite{Peik1,Campbell2}, and the crystal lattice nuclear clock, based on multiple $^{229}$Th ions embedded in a crystal lattice environment \cite{Rellergert,Kazakov1}. Trapping and laser cooling of $^{229}$Th$^{3+}$ ions as a prerequisite for a single ion nuclear clock was achieved in 2011 \cite{Campbell2011}. A recent review on the status of the nuclear clock development is provided in Ref.~\cite{Wense2018}. The proposal for the development of a nuclear clock of unprecedented accuracy has triggered a multitude of experimental efforts aiming to determine the nuclear parameters of $^{229\text{m}}$Th to higher precision. Especially the energy is of importance in this context, as it determines the required laser technology for nuclear excitation and reduces the time for laser-based scanning when searching for the nuclear excitation. The direct detection of light emitted in the isomeric decay would significantly help to pin down the energy value.\\[0.2cm] 
Early experiments searching for the direct detection of the isomeric decay failed to observe any signal. This can partly be explained by a correction of the measured transition energy to 7.6~eV in 2007 \cite{Beck1}, later slightly shifted to 7.8~eV \cite{Beck2}. While until today still no conclusive detection of light emitted in the isomer's decay was achieved \cite{Jeet,Yamaguchi,Wense2,Stellmer}, a 2016 measurement succeeded in the observation of electrons as emitted in the isomer's internal conversion (IC) decay channel \cite{Wense1}. In the IC decay the nucleus transfers its energy to the electronic shell, leading to the subsequent ejection of an electron. The observation of the IC electrons allowed for a first lifetime measurement of $^{229\text{m}}$Th contained in neutral, surface bound atoms, which was found to be about 10 $\mu$s \cite{Seiferle3}, in agreement with theoretical expectations \cite{Strizhov,Karpeshin1,Tkalya4}.\\[0.2cm]
In 2018, collinear laser-spectroscopy of the electron shell of $^{229\text{m}}$Th$^{2+}$ compared to $^{229}$Th$^{2+}$ allowed to measure the isomer-induced hyperfine shift \cite{Thielking}. Based on this measurement, the magnetic dipole and electric quadrupole moment as well as the mean-squared charge radius of the nucleus in the metastable state could be determined. In 2019, a successful excitation of the $^{229}$Th 29~keV nuclear state via synchrotron radiation was reported \cite{Masuda}. The 29~keV state decays with a probability of 90\% to the isomeric state, thereby allowing for a new way of efficient population of $^{229\text{m}}$Th. Also the $^{229\text{m}}$Th energy was constrained to a value of $8.28\pm0.17$~eV with higher precision via spectroscopy of the IC electrons emitted in the isomeric decay \cite{Seiferle4}.\\[0.2cm]
Based on the direct detection of the isomer's IC decay channel \cite{Wense1} in combination with the short isomeric lifetime \cite{Seiferle3} and the improved energy value \cite{Seiferle4}, new concepts for direct nuclear laser excitation were presented \cite{Wense3,Wense4,Wense2019}. These make use of IC electron detection for probing the isomer's successful excitation, which has the advantage that the detection can be triggered in coincidence with the laser pulses. In this way a large signal-to-background ratio and short times required for laser-based scanning are expected. These experiments offer the potential to determine the isomeric energy value to a precision corresponding to the bandwidth of the laser light used for scanning, thereby providing the basis for the development of a nuclear clock \cite{Wense2019}. To support these upcoming experimental studies with a solid theoretical framework, a comprehensive analytical consideration of direct nuclear laser excitation using the density-matrix formalism is presented. The theory of direct nuclear laser excitation is identical to the theory of atomic laser excitation, with only three differences:\\[0.2cm]
\textbf{(i)} In the interaction Hamiltonian, the electronic current density operator is exchanged by the nuclear current density operator.\\[0.2cm]
\textbf{(ii)} For the sake of generality the theory is considered for arbitrary multipole order. This allows the discussion of nuclei other than $^{229}$Th.\\[0.2cm]
\textbf{(iii)} Besides the radiative decay channel, also different non-radiative decay channels have to be considered. The most important one is the IC decay.\\[0.2cm]
Previous related studies are Refs. \cite{Buervenich,Palffy,Palffy2,Dzyublik}, where the case of intense X-ray laser fields driving transitions to low-lying nuclear levels was discussed. More advanced schemes, which employ two fields to achieve nuclear coherent population transfer, were also presented \cite{Liao1,Liao3}. The theory of $^{229\text{m}}$Th laser excitation was discussed in Refs.~\cite{Kazakov1,Kazakov2} and, for the two-photon case, in Ref.~\cite{Romanenko}.\\[0.2cm]
This paper is structured as follows: In Sec.~\ref{sec2} the optical Bloch equations for the simplest case of a nuclear two-level system are derived. In Sec.~\ref{secRabi} the explicit form of the Rabi frequency will be obtained as an input parameter for the optical Bloch equations. Analytic solutions of the two-level optical Bloch equations for different cases will be presented in Sec.~\ref{sec5}. The case of a nuclear two-level system consisting of sub-states will be discussed in Sec.~\ref{sec3} for free atoms and ions as well as for the solid-state environment. The resulting multi-state optical Bloch equations are analytically solved for laser fields of moderate intensity and a linewidth significantly broader than the one of the nuclear transition in Sec.~\ref{sec4}. In Sec.~\ref{sec7} the theory of nuclear two-photon excitation will be considered. Finally, the special cases of $^{229\text{m}}$Th and $^{235\text{m}}$U will be discussed in Sec.~\ref{sec8} and Sec.~\ref{sec9}, respectively. 

\section{\label{sec2}The optical Bloch equations for a two-level system}
\begin{figure}
\begin{center}
  \includegraphics[height=6cm]{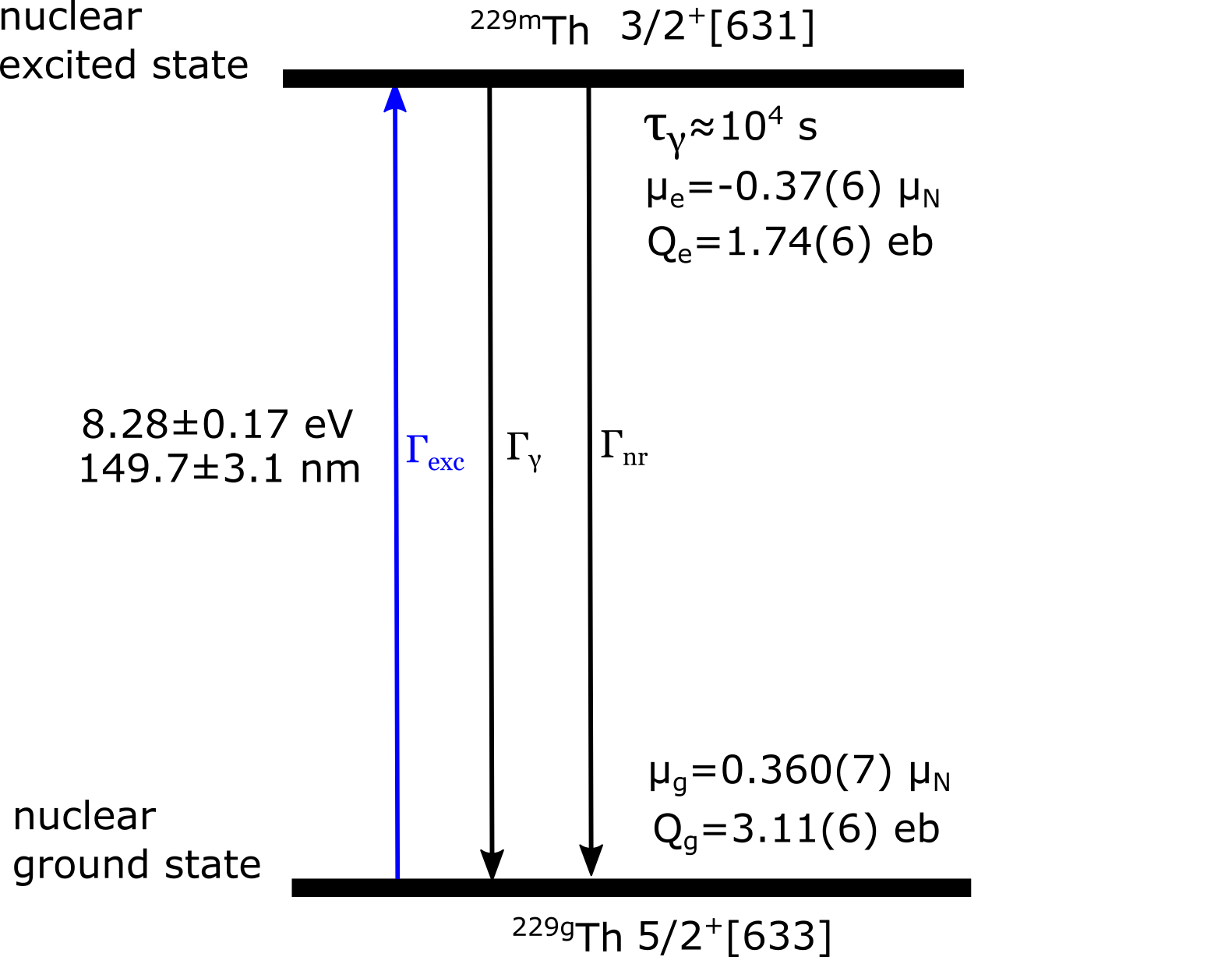}
  \caption{Example for a nuclear two-level system: the closely spaced energy doublet of ground and metastable excited state at approximately 8.3~eV in $^{229}$Th.}
  \label{twolevel}
	\end{center}
\end{figure}

\noindent Population transfer due to coherent laser excitation is generally described by the optical Bloch equations (see e.g. Refs.~\cite{Scully,Steck}), which model the time development of the density operator $\hro$ determining the population probabilities of the individual states. As an example for the simplified case of a two-level system, the transition between the ground- and excited state of $^{229}$Th is shown in Fig.~\ref{twolevel}. The starting point for the derivation of the optical Bloch equations for nuclear excitation is the quantum optical master equation in Lindblad form given in the Schrödinger picture as \cite{Scully,Lindblad}
\begin{equation}
\colorbox{lightgray}{$
\begin{aligned}
\frac{\partial\hro(t)}{\partial t}=-\frac{i}{\hbar}\left[\H(t),\hro(t)\right]+\LL[\hro].
\end{aligned}
$}
\label{master}
\end{equation}
This equation, without the term $\LL[\hro]$, would be identical to the von-Neumann equation, which describes the time evolution of the density operator $\hro$ without energy dissipation. The additional term, $\LL[\hro]$, is the Lindblad superoperator (an operator with the density operator as an argument), which describes relaxation processes like spontaneous decay. $\H(t)=\H_N+\H_I(t)$ denotes the total Hamiltonian, a sum of the nuclear Hamiltonian $\H_N$ (including the influence of external constant electric or magnetic fields created by the environment), and the interaction Hamiltonian $\H_I(t)$ describing the coupling of the nucleus to the external laser field (treated as classical). The density operator $\hro$ can be represented as
\begin{equation}
\hro(t)=\sum_{i,j} \rho_{ij}|i\rangle\langle j|,
\label{densop}
\end{equation}
where $|i\rangle$ and $|j\rangle$ are eigenstates of $\H_N$, defined as
\begin{equation}
\H_N = \hbar \sum_{i} \omega_{i}|i\rangle\langle i|.
\label{nuclHams1}
\end{equation}
Here $\hbar \omega_i$ is the energy of the state $|i\rangle$. In this section the simplest case of a two-level system consisting only of ground and excited state, denoted as $|g\rangle$ and $|e\rangle$, is considered. In this case, the density operator $\hat{\rho}(t)$, defined in Eq.~(\ref{densop}), takes the explicit form 
\begin{equation}
\begin{aligned}
\hat{\rho}(t)=&\rho_{ee}\vert e\rangle\langle e\vert+\rho_{ge}\vert g\rangle\langle e\vert+\rho_{eg}\vert e\rangle\langle g\vert+\rho_{gg}\vert g\rangle\langle g\vert,
\end{aligned}
\label{densop2}
\end{equation}
where $\rho_{ee}$ and $\rho_{gg}$ denote the population probabilities of the ground and excited nuclear state, respectively, and $\rho_{ge}=\rho_{eg}^*$ are the so-called coherences. Further, the nuclear Hamiltonian [Eq.~(\ref{nuclHams1})] takes the form
\begin{equation}
\H_N = \hbar \omega_{e}|e\rangle\langle e|+
\hbar \omega_{g}|g\rangle\langle g|.
\label{nuclHams}
\end{equation}
The interaction Hamiltonian $\H_I(t)$ of the nucleus with a single-mode excitation laser in Coulomb gauge reads \cite{Palffy,Ring}
\begin{multline}
\hat{H}_I(t)=- \int \hvj \cdot \vA(\r,t) d^3 \r\\
=- \int \hvj \cdot \left[\vA_0 e^{i(\k_\ell\cdot \r-\omega_\ell t-\phi(t))}+\text{c.c.}\right] d^3 \r,
\label{hamiltonian1}
\end{multline}
where $\hvj$ is the operator of the nuclear current density, $\vA(\r,t)$ denotes the vector potential of the laser field defined as
\begin{equation}
\vec{A}(\vec{r},t)=\vec{A}_0 e^{i(\vec{k}_\ell\cdot\vec{r}-\omega_\ell t-\phi(t))}+\vec{A}_0^* e^{-i(\vec{k}_\ell\cdot\vec{r}-\omega_\ell t-\phi(t))},
\end{equation}
with complex amplitude $\vA_0$, angular frequency $\omega_\ell$, wave vector $\k_\ell$ and fluctuating phase $\phi(t)$. Phase fluctuations of the laser field used for nuclear excitation lead to larger decay rates of the coherences. Further, they also influence the spectrum of the laser field \cite{Riehle}. In this way, broad-band laser sources lead to short coherence times when used to excite the optical transitions (see Appendix~A for a discussion).\\[0.2cm]
Assuming linearly polarized light, the electric field $\vec{E}(\vec{r},t)$ is obtained from $\vec{E}=-\partial\vec{A}/\partial t$ to be
\begin{equation}
\label{Evec}
\begin{aligned}
\vec{E}(\vec{r},t)&= \frac{\vec{E}_0}{i}\left( e^{-i(\vec{k}_\ell\cdot\vec{r}-\omega_\ell t-\phi(t))}- e^{i(\vec{k}_\ell\cdot\vec{r}-\omega_\ell t-\phi(t))}\right)\\
&=2\vec{E}_{0} \sin\left(-(\vec{k}_\ell\cdot\vec{r}-\omega_\ell t-\phi(t))\right),
\end{aligned}
\end{equation}
where $\vec{E}_0=\omega_\ell\vec{A}_0$ was used. A similar equation for the magnetic field $\vec{B}(\vec{r},t)$ is obtained using $\vec{B}(\vec{r},t)=\vec{E}(\vec{r},t)/c$ and defining $\vec{B}_0=\vec{E}_0/c$.\\[0.2cm]
Eq.~(\ref{hamiltonian1}) can be rewritten to give
\begin{equation}
\hat{H}_I(t)=\hat{H}_I^0 e^{-i\left(\omega_\ell t+\phi(t)\right)}
            +\hat{H}_I^{0\dagger}e^{i\left(\omega_\ell t+\phi(t)\right)},
\label{hamiltonian2}
\end{equation}
where the time independent interaction Hamiltonian $\hat{H}_I^0$ was introduced as
\begin{equation}
\hat{H}_I^0=- \int \hvj \cdot\vA_0 e^{i\k_\ell\cdot\r} d^3 \r.
\end{equation}
For the following the interaction matrix elements $U_{ge}$ and $U_{eg}$ as well as $V_{ge}$ and $V_{eg}$ are defined as
\begin{equation}
\label{Vdef}
\begin{aligned}
U_{ge}=\frac{\langle g\vert \hat{H}_I^{0}\vert e\rangle}{\hbar},&\quad U_{eg}=\frac{\langle e\vert \hat{H}_I^{0\dagger}\vert g\rangle}{\hbar},\\
V_{ge}=\frac{\langle g\vert \hat{H}_I^{0\dagger}\vert e\rangle}{\hbar},&\quad V_{eg}=\frac{\langle e\vert \hat{H}_I^{0}\vert g\rangle}{\hbar}.\\
\end{aligned}
\end{equation}
In this way, $U_{eg}=U_{ge}^*$ and $V_{eg}=V_{ge}^*$ hold. Further, it is assumed that the laser intensity is sufficiently low to neglect the diagonal terms of the interaction Hamiltonian, which would otherwise lead to higher order processes like multi-photon excitation, discussed in Sec.~\ref{sec7} and Ref.~\cite{Romanenko}. Therefore one obtains
\begin{equation}
\label{hamiltonian3b}
\begin{aligned}
&\hat{H}_I^{0}&=\hbar \left(U_{ge}|g\rangle\langle e|+ V_{eg}|e\rangle\langle g|\right), \\
&\hat{H}_I^{0\dagger}&=\hbar \left(U_{eg}|e\rangle\langle g|+ V_{ge}|g\rangle\langle e|\right).
\end{aligned}
\end{equation}
\\[0.2cm]
Based on the definition of the density operator in Eq.~(\ref{densop2}), the population probability of the excited state $|e\rangle$ is obtained as
\begin{equation}
\rho_{ee}=\langle e\vert\hat{\rho}(t)\vert e\rangle.
\label{matrixelements}
\end{equation}
For an explicit derivation of the optical Bloch equations, first the von-Neumann term of Eq.~(\ref{master}) is considered. This term transforms to
\begin{equation}
\begin{aligned}
&-\frac{i}{\hbar}\left[\hat{H}(t),\hat{\rho}(t)\right]=-\frac{i}{\hbar}\left[\hat{H}_{N}+\hat{H}_{I}(t),\hat{\rho}(t)\right]\\
=&-\frac{i}{\hbar}\left(\hat{H}_N\hat{\rho}(t)-\hat{\rho}(t)\hat{H}_N+\hat{H}_I(t)\hat{\rho}(t)-\hat{\rho}(t)\hat{H}_I(t)\right).\\
\end{aligned}
\label{vonneumann}
\end{equation}
In the following, only the $\dot{\rho}_{ee}$ component of the master equation Eq.~(\ref{master}) will be discussed. All remaining components can be obtained by similar calculations. For this purpose the unity operator \(\mathds{\hat{1}}\) is inserted between each operator product of Eq.~(\ref{vonneumann}) as
\begin{equation}
\mathds{\hat{1}}=\vert g\rangle\langle g\vert+\vert e\rangle\langle e\vert.
\end{equation}
Then, for the von-Neumann term one obtains
\begin{equation}
\begin{aligned}
 &-\frac{i}{\hbar} \langle e\vert\left[\hat{H}(t),\hat{\rho}(t) \right]\vert e\rangle \\
=&-\frac{i}{\hbar}\Big[\rho_{ge}\langle e\vert \hat{H}_I(t)\vert g\rangle-\rho_{eg}\langle g\vert \hat{H}_I(t)\vert e\rangle \Big]\\
=&-i\Big[\left(V_{eg}\rho_{ge}-U_{ge}\rho_{eg}\right)e^{-i(\omega_\ell t+\phi(t))}\\
&\quad+\left(U_{eg}\rho_{ge}-V_{ge}\rho_{eg}\right)e^{i(\omega_\ell t+\phi(t))}\Big].
\end{aligned}
\end{equation}
Here $\hat{H}_I(t)$ was inserted according to Eq.~(\ref{hamiltonian2}) and $\hat{H}_I^{0}$ as well as $\hat{H}_I^{0\dagger}$ were used as defined in Eq.~(\ref{hamiltonian3b}).\\[0.2cm]
Next the Lindblad superoperator $\mathcal{L}[\hat{\rho}]$ is considered, which is the second term on the right-hand side of Eq.~(\ref{master}). This term models various relaxation processes due to interaction with the environment. In this section only spontaneous decay of the excited state $|e\rangle$ into the ground state $|g\rangle$ is taken into consideration, described by the following Lindblad operator \cite{Scully,Lindblad}
\begin{equation}
\LL[\hat{\rho}]=\Gamma\left(\hat{\sigma}_{ge}\hat{\rho}\hat{\sigma}_{ge}^\dagger-\frac{1}{2}\hat{\rho}\hat{\sigma}_{ge}^\dagger\hat{\sigma}_{ge}-\frac{1}{2}\hat{\sigma}_{ge}^\dagger\hat{\sigma}_{ge}\hat{\rho}\right),
\label{lindbladsp}
\end{equation}
where $\Gamma=\Gamma_\gamma+\Gamma_{nr}$ denotes the total decay rate of the excited state, including radiative as well as non-radiative decay channels (e.g., IC) with rates $\Gamma_\gamma$ and $\Gamma_{nr}$, respectively.
Importantly, here it is assumed that all decay channels are populating the same ground state. This allows us to treat the system as an effective two-level system. In reality, this condition is usually not satisfied, which requires to consider multi-level systems as discussed in Sec.~\ref{sec3}. The remaining operators are defined as $\hat{\sigma}_{ge}=\vert g \rangle\langle e \vert$ and $\hat{\sigma}_{ge}^\dagger=\hat{\sigma}_{eg}=\vert e \rangle\langle g \vert$. Using the definition in Eq.~(\ref{lindbladsp}), a straightforward calculation reveals
\begin{equation}
\langle e\vert\ \mathcal{L}[\hat{\rho}]\ \vert e\rangle=-\rho_{ee}\Gamma.
\end{equation}
Performing the same calculation also for the remaining matrix elements $\rho_{gg}$ and $\rho_{ge}=\rho_{eg}^*$ leads to the optical Bloch equations for a two-level system including relaxation in the Schrödinger picture:\\
\begin{equation}
\begin{aligned}
\dot{\rho}_{ee}=-\dot{\rho}_{gg}=&-i\Big[V_{eg}\rho_{ge}-U_{ge}\rho_{eg}\Big]e^{-i(\omega_\ell t+\phi(t))}\\ 
&-i\Big[U_{eg}\rho_{ge}- V_{ge}\rho_{eg}\Big]e^{i(\omega_\ell t+\phi(t))}-\rho_{ee}\Gamma,\\
\dot{\rho}_{ge}=\dot{\rho}_{eg}^*=&-iV_{ge}\Big[\rho_{ee}-\rho_{gg}\Big]e^{i(\omega_\ell t+\phi(t))}\\
&-iU_{ge}\Big[\rho_{ee}-\rho_{gg}\Big]e^{-i(\omega_\ell t+\phi(t))}\\
&+\rho_{ge}\Big[i\omega_0-\Gamma/2\Big].\\
\end{aligned}
\label{bloch}
\end{equation}

\noindent Here $\omega_0=\omega_e-\omega_g$ was introduced. In order to avoid the fast oscillating time dependent terms and also the huge term proportional to $\omega_0$, it is convenient to introduce a transformation into the rotating frame:
\begin{equation}
\label{rotframe}
\begin{aligned}
\tilde{\rho}_{ee}&=\rho_{ee},\quad &&\tilde{\rho}_{gg}=\rho_{gg} \\
\tilde{\rho}_{ge}&=\rho_{ge}e^{-i(\omega_\ell t+\phi(t))}, \quad &&\tilde{\rho}_{eg}=\rho_{eg}e^{i(\omega_\ell t+\phi(t))}.\\
\end{aligned}
\end{equation}
After this transformation, the rotating wave approximation is applied, which is valid close to resonance and consists in neglecting the fast oscillating terms proportional to $e^{-i2\omega_\ell}$ and $e^{i2\omega_\ell}$. Then the optical Bloch equations take the following form:
\begin{equation}
\begin{aligned}
\dot{\tilde{\rho}}_{ee}&=-\dot{\tilde{\rho}}_{gg}=-i\Big[V_{eg}\tilde{\rho}_{ge} - V_{ge} \tilde{\rho}_{eg} \Big]-\tilde{\rho}_{ee}\Gamma,\\
\dot{\tilde{\rho}}_{ge}&=\dot{\tilde{\rho}}_{eg}^*=-iV_{ge}\Big[\tilde{\rho}_{ee}-\tilde{\rho}_{gg}\Big]-\tilde{\rho}_{ge}\Big[i\Delta\omega+i\dot{\phi}+\Gamma/2\Big].\\
\end{aligned}
\label{bloch2}
\end{equation}
Here the detuning $\Delta\omega=\omega_\ell-\omega_0$ between the angular frequency of the laser light with respect to the transition angular frequency was introduced. In case the laser spectrum possesses a Lorentzian shape with a full width at half maximum (FWHM) of $\Gamma_\ell$, averaging the equation for $\dot{\tilde{\rho}}_{ge}$ in Eq.~(\ref{bloch2}) over the fluctuating laser phase $\phi(t)$ (see Appendix A for details) leads to the optical Bloch equations in the form:
\begin{equation}
\begin{aligned}
\dot{\rho}_{ee}&=-\dot{\rho}_{gg}=-i\Big[V_{eg}\rho_{ge} - V_{ge}\rho_{eg}\Big]-\rho_{ee}\Gamma,\\
\dot{\rho}_{ge}&=\dot{\rho}_{eg}^*=-iV_{ge}\Big[\rho_{ee}-\rho_{gg}\Big]-\rho_{ge}\Big[i\Delta\omega+\tilde{\Gamma}\Big].\\
\end{aligned}
\label{bloch3}
\end{equation}
The tilde-signs of the density matrix elements were dropped here and the decay rate of the coherences $\tilde{\Gamma}$ was defined as
\begin{equation}
\label{Gammatilde}
\tilde{\Gamma}=\frac{\Gamma+\Gamma_\ell}{2}.
\end{equation}
Therefore, a short coherence time of the laser light introduces an additional decay rate of the coherences \cite{Steck}. In a last transformation one can substitute
\begin{equation}
\rho_{ge}=\frac{\vert V_{eg} \vert}{V_{eg}}\bar{\rho}_{ge} 
\end{equation}
to obtain (after dropping the bar signs) the optical Bloch equations for a two-level system in the final form \cite{Steck}:
\begin{equation}
\label{twolevelbloch}
\colorbox{lightgray}{$
\begin{aligned}
\dot{\rho}_{ee}&=-\dot{\rho}_{gg}=-i\frac{\Omega_{eg}}{2}\Big[\rho_{ge}-\rho_{eg}\Big]-\rho_{ee}\Gamma,\\
\dot{\rho}_{ge}&=\dot{\rho}_{eg}^*=-i\frac{\Omega_{eg}}{2}\Big[\rho_{ee}-\rho_{gg}\Big]-\rho_{ge}(i\Delta\omega+\tilde{\Gamma}).
\end{aligned}
$}
\end{equation}
Here the Rabi frequency $\Omega_{eg}$ is defined as 
\begin{equation}
\label{Rabi}
\Omega_{eg}=2\vert V_{eg}\vert=\frac{2\vert\langle e\vert\hat{H}_I^0\vert g\rangle\vert}{\hbar}.
\end{equation}
The explicit form of $\Omega_{eg}$ will be derived in the following section. The optical Bloch equations for a two-level system without magnetic sub-states, Eq.~(\ref{twolevelbloch}), can be solved analytically \cite{Noh}. Analytic solutions for three different cases will be discussed in Sec.~\ref{sec5}: for zero detuning ($\Delta\omega=0$), for the low-saturation limit ($\rho_{gg}\gg\rho_{ee}$ and $\Gamma_\ell\gg\Gamma$) and for the steady-state case ($\dot{\rho}_{ee}=\dot{\rho}_{gg}=\dot{\rho}_{ge}=0$). The optical Bloch equations for the more general cases of nuclear multi-level systems under different conditions (nuclei in individual atoms or ions and in a solid-state environment) will be discussed in Sec.~\ref{sec3}.


\section{\label{secRabi}Calculating the Rabi frequency for nuclear transitions}
In order to make use of Eq.~(\ref{twolevelbloch}), the Rabi frequency $\Omega_{eg}$ has to be determined explicitly. Considering Eq.~(\ref{Rabi}), this requires the evaluation of the absolute value of the interaction matrix element $\lvert\langle e\vert\hat{H}_I^0\vert g\rangle\rvert$. The interaction Hamiltonian $\hat{H}_I^0$ can be expressed as a sum of terms consisting of irreducible tensors of multipolarity $\lambda L$ with magnetic index $\sigma$:
\begin{equation}
\label{summation}
\hat{H}_I^0=\sum_{\lambda L,\sigma}\hat{H}_{I\ \lambda L}^{0\ \sigma}.
\end{equation}
Here $\lambda$ stands for the electric ($E$) or magnetic ($M$) case and defines the parity of the field. $L$ and $\sigma$ denote the corresponding angular momentum and magnetic quantum number, respectively. Eq.~(\ref{summation}) corresponds to a decomposition of the interaction Hamiltonian into operators that represent radiation of different multipole orders of either electric or magnetic multipolarity. For the calculation of the Rabi frequency $\Omega_{eg}$, Eq.~(\ref{summation}) is inserted into Eq.~(\ref{Rabi}) and interference of different multipole channels is neglected. In this case one can bring the summation over different $\lambda L$ outside the absolute value in Eq.~(\ref{Rabi}), leading to
\begin{equation}
\Omega_{eg}=\frac{2}{\hbar}\sum_{\lambda L,\sigma} \vert\langle e\vert\hat{H}_{I\ \lambda L}^{0\ \sigma}\vert g\rangle\vert.
\end{equation}
For all following considerations it is assumed that one particular multipolarity $\lambda L$ dominates over the others. This allows to restrict the discussion to a pair of defined quantum numbers $\lambda L$ and $\sigma$. In a physical interpretation this equals a restriction to photons corresponding to either magnetic or electric multipole radiation with angular momentum $L$ and magnetic quantum number $\sigma$. Therefore one can drop the summation, which leads to
\begin{equation}
\label{Omegaeg}
\Omega_{eg}=\frac{2 \vert\langle e\vert\hat{H}_{I\ \lambda L}^{0\ \sigma}\vert g\rangle\vert}{\hbar}.
\end{equation}
The interaction matrix element $\langle e\vert \hat{H}_{I\ \lambda L}^{0\ \sigma} \vert g\rangle$ for magnetic multipole radiation ($\lambda=M$) can be explicitly expressed in terms of the magnetic multipole operator $\hat{\mathcal{M}}_{L\sigma}$ as \cite{Palffy}
\begin{equation}\label{RabiFreq1}
\begin{aligned}
&\left| \bra{e} \hat{H}_{I\ M L}^{0\ \sigma} \ket{g} \right|\\
=\ &B_0 \sqrt{2 \pi} \sqrt{\frac{L+1}{L}}\frac{k^{L-1}}{(2L+1)!!} \left| \bra{e} \hat{\mathcal{M}}_{L\sigma} \ket{g} \right|\;.
\end{aligned}
\end{equation}
Note, that compared to Ref.~\cite{Palffy} the equation was divided by $\sqrt{2L+1}$ in order to take account for a random polarization of the light used for excitation. The wave vector $k_\ell$ of the laser light is assumed to be directed along the quantization axis of the system leading to equally probable values of $\sigma$. Here, $B_{0}$ denotes the amplitude of the classical magnetic field that drives the transition and $k=\omega_0/c$ is the wavenumber corresponding to the nuclear transition. Similarly, for electric multipole radiation ($\lambda=E$) one obtains the interaction matrix element in terms of the electric multipole operator $\hat{\mathcal{Q}}_{L\sigma}$ as \cite{Palffy}
\begin{equation}\label{RabiFreq2}
\begin{aligned}
&\left| \bra{e} \hat{H}_{I\ E L}^{0\ \sigma} \ket{g} \right|\\
=\ &E_0 \sqrt{2 \pi} \sqrt{\frac{L+1}{L}}\frac{k^{L-1}}{(2L+1)!!} \left| \bra{e} \hat{\mathcal{Q}}_{L\sigma} \ket{g} \right|\;,
\end{aligned}
\end{equation}
with $E_{0}=cB_0$ being defined via the classical electric driving field $\vec{E}$ as in Eq.~(\ref{Evec}). Note, that the explicit derivation of these expressions is involved and can be found, for example, in Refs.~\cite{Liao2,Palffy,Ring}. The magnetic multipole operator was introduced in agreement with Ref.~\cite{Ring} as
\begin{equation}\label{Momentum1}
\hat{\mathcal{M}}_{L\sigma}=\frac{1}{L+1}\int d^3r \left[ \vec{r}\times\vec{j} \right]\nabla \left( r^L Y_{L\sigma} \right)\;
\end{equation}
and the electric multipole operator as \cite{Ring}
\begin{equation}\label{Momentum2}
\hat{\mathcal{Q}}_{L\sigma}=\int d^3r \rho\ r^L Y_{L\sigma},
\end{equation}
where $\vec{j}$ and $\rho$ denote the nuclear current density and charge distribution, respectively, and $Y_{L\sigma}$ are the spherical harmonics. On the other hand, the radiative decay rate between the states $\ket{e}$ and $\ket{g}$ can be expressed as~\cite{Ring}
\begin{equation}\label{magneticdecay}
\Gamma_{ge}^\gamma=
\frac{2\mu_0}{\hbar}
\frac{L+1}{L[(2L+1)!!]^2}
k^{2L+1}
\left| \langle g \vert \hat{\mathcal{M}}_{L\sigma}\vert e \rangle \right| ^2 \;
\end{equation}
in case of magnetic multipole radiation, and as 
\begin{equation}\label{electricdecay}
\Gamma_{ge}^\gamma=
\frac{2}{\hbar\epsilon_0}
\frac{L+1}{L[(2L+1)!!]^2}
k^{2L+1}
\left|\langle g \vert \hat{\mathcal{Q}}_{L\sigma}\vert e \rangle \right| ^2 \;
\end{equation}
in case of electric multipole radiation. By using that
\begin{equation}
\begin{aligned}
\langle g \vert \hat{\mathcal{M}}_{L\sigma}\vert e \rangle=& \langle e \vert \hat{\mathcal{M}}_{L\sigma}\vert g \rangle,\\
\langle g \vert \hat{\mathcal{Q}}_{L\sigma}\vert e \rangle=& \langle e \vert \hat{\mathcal{Q}}_{L\sigma}\vert g \rangle\\
\end{aligned}
\end{equation}
and inserting Eq.~(\ref{magneticdecay}) into Eq.~(\ref{RabiFreq1}), after application of $B_0=E_0/c$ and $c=1/\sqrt{\mu_0\epsilon_0}$ the interaction matrix element for magnetic radiation is obtained as
\begin{equation}
\left| \bra{e} \hat{H}_{I\ M L}^{0\ \sigma} \ket{g} \right|=\sqrt{\frac{E_0^2\pi\epsilon_0\hbar \Gamma_{ge}^\gamma}{k^3}}.
\end{equation}
The same equation holds also for the electric radiation type as can be obtained by inserting Eq.~(\ref{electricdecay}) into Eq.~(\ref{RabiFreq2}). Therefore we can write a generalized expression as
\begin{equation}
\label{nonreduced0}
\left| \bra{e} \hat{H}_{I\ \lambda L}^{0\ \sigma} \ket{g} \right|=\sqrt{\frac{E_0^2\pi\epsilon_0\hbar \Gamma_{ge}^\gamma}{k^3}}.
\end{equation}
Note, that different parity selection rules for the cases $\lambda=E$ and $\lambda=M$ apply.\\[0.2cm]
In a next step, the interaction matrix element [Eq.~(\ref{nonreduced0})] will be expressed as a function of the laser intensity $I_\ell$. The total time-averaged power density per unit square (intensity) of the electromagnetic wave is given as
\begin{equation}
I_\ell=c\frac{\epsilon_{0}\overline{\vert \vec{E}(\vec{r},t)\vert^{2}}}{2}+c\frac{\overline{\vert \vec{B}(\vec{r},t)\vert^{2}}}{2\mu_{0}}.
\end{equation}
By applying $\vec{B}(\vec{r},t)=\vec{E}(\vec{r},t)/c$, inserting Eq.~(\ref{Evec}) for $\vec{E}(\vec{r},t)$ and using $\overline{\sin^{2}(-(\vec{k}_\ell\cdot\vec{r}-\omega_\ell t))}=1/2$, one obtains
\begin{equation}
\label{Itot1}
I_\ell=c\epsilon_{0}\overline{\vert \vec{E}(\vec{r},t)\vert^{2}}=2c\epsilon_{0} E_{0}^{2},
\end{equation}
The resulting expression $E_0^2=I_\ell/(2c\epsilon_{0})$ together with $k=\omega_0/c$ is inserted into Eq.~(\ref{nonreduced0}), which transforms to
\begin{equation}
\label{nonreduced1}
\left| \bra{e} \hat{H}_{I\ \lambda L}^{0\ \sigma} \ket{g} \right|=\sqrt{\frac{\pi\hbar c^2 I_\ell \Gamma_{ge}^\gamma}{2\omega_0^3}}.
\end{equation}
Now inserting Eq.~(\ref{nonreduced1}) for the matrix element of the interaction Hamiltonian into Eq.~(\ref{Omegaeg}) one obtains for the Rabi frequency
\begin{equation}
\label{Omega01}
\colorbox{lightgray}{$
\begin{aligned}
\Omega_{eg}=\sqrt{\frac{2\pi c^2 I_\ell \Gamma_{ge}^\gamma}{\hbar\omega_0^3}}.
\end{aligned}
$}
\end{equation}
The Rabi frequency contains the radiative decay rate $\Gamma_{eg}^\gamma$ between the excited state and the ground state. The explicit form of $\Gamma_{eg}^\gamma$ requires knowledge about the physical situation and the particular pair of sub-states under consideration. It will be derived in Sec.~\ref{sec3}, where the nuclear hyperfine splitting (HFS) is taken into consideration. It is found that in case of isolated atoms or ions (e.g., $^{229}$Th$^{3+}$ ions in a Paul trap), $\Gamma_{eg}^\gamma$ can be expressed in terms of the total radiative decay rate $\Gamma_\gamma$ by means of Eq.~(\ref{radiativedecayrate1}). Opposed to that, in a solid-state environment Eq.~(\ref{gammam}) has to be used.

\section{\label{sec5}Solutions of the two-level optical Bloch equations}
In this section solutions of the two-level optical Bloch equations given in Eq.~(\ref{twolevelbloch}) for different scenarios are presented. Note that Eq.~(\ref{twolevelbloch}) possesses a complete analytical solution, which was derived in Ref.~\cite{Noh}. In the following, however, only the three most important cases will be considered: (1) The case of zero detuning of the laser light with respect to the resonance ($\Delta\omega=0$), (2) The low-saturation limit, which assumes the laser intensity to be low, such that the excited state is far less populated than the ground state ($\rho_{gg}\gg\rho_{ee}$) and (3) the steady-state solution obtained after long interrogation times. Each solution will be discussed individually.

\subsection{\label{sec3_1}Solution for zero detuning}
The solution of the optical Bloch equations, Eq.~(\ref{twolevelbloch}), for zero detuning of the laser light with respect to the resonance ($\Delta\omega=0$) and for the initial condition $\rho_{ee}(t=0)=0$ is known as Torrey's solution \cite{Torrey}. Its derivation can be found, e.g., in Refs.~\cite{Steck,Noh}. Here, merely the result is quoted as\footnote{To obtain Torrey's solution in the form of Eq.~(\ref{Torrey}), $\rho_{ee}=1/2(1-w)$ was used with $w$ taken from Eq.~(16) of Ref.~\cite{Noh}.}\\
\begin{equation}
\label{Torrey}
\colorbox{lightgray}{$
\begin{aligned}
\rho_{ee}=&\frac{\Omega_{eg}^2}{2\left(\Gamma\tilde{\Gamma}+\Omega_{eg}^2\right)}\times\\ &\left[1-e^{-\frac{1}{2}(\Gamma+\tilde{\Gamma})t}\left(\cos(\lambda t)+\frac{\Gamma+\tilde{\Gamma}}{2\lambda}\sin(\lambda t)\right)\right],
\end{aligned}
$}
\end{equation}
where $\lambda$ was defined as $\lambda=|\Omega_{eg}^2-(\tilde{\Gamma}-\Gamma)^2/4|^{1/2}$. Eq.~(\ref{Torrey}) is only valid for $\lvert\tilde{\Gamma}-\Gamma\rvert/2<\Omega_{eg}$.
In case of $\lvert\tilde{\Gamma}-\Gamma\rvert/2>\Omega_{eg}$, the $\sin$ and $\cos$ functions have to be exchanged by $\sinh$ and $\cosh$, respectively.\\[0.2cm]
Eq.~(\ref{Torrey}) remains valid also if the bandwidth of the laser light $\Gamma_\ell$ is significantly broader than the linewidth of the transition $\Gamma$, as the definition of the decay rate of the coherences $\tilde{\Gamma}$ in Eq.~(\ref{Gammatilde}) takes that into account.\\[0.2cm]
The equation describes Rabi oscillations of the nuclear two-level system if the Rabi frequency $\Omega_{eg}$, obtained in Eq.~(\ref{Omega01}) is larger than the decay rate of the coherences $\tilde{\Gamma}$. For the development of a single-ion nuclear clock, Rabi oscillations are important, as they allow to apply the Ramsey-interrogation scheme in the clock concept, which in turn improves the clock stability to a value corresponding to the Allan deviation limited by the quantum-projection noise \cite{Ludlow}. Using Eq.~(\ref{Gammatilde}) for the definition of $\tilde{\Gamma}$, the condition $\Omega_{eg}\gg\tilde{\Gamma}$ is fulfilled if $\Omega_{eg}\gg\text{max}(\Gamma/2,\Gamma_\ell/2)$ holds. With Eq.~(\ref{Omega01}) for the Rabi frequency, the condition for observing Rabi oscillations can be expressed as
\begin{equation}
\label{Rabiuneq}
\sqrt{\frac{2\pi c^2 I_\ell \Gamma_{ge}^\gamma}{\hbar\omega_0^3}} \gg \text{max}(\Gamma/2,\Gamma_\ell/2).
\end{equation}
In case that the bandwidth of the laser light is broader than the nuclear transition linewidth ($\Gamma_\ell\gg\Gamma$), Eq.~(\ref{Rabiuneq}) leads to the condition for the laser intensity required to drive Rabi oscillations as 
\begin{equation}
\label{I1}
I_\ell\gg \frac{\hbar\omega_0^3\Gamma_\ell^2}{8\pi c^2\Gamma_{ge}^\gamma}.
\end{equation}
On the other hand, for $\Gamma\gg\Gamma_\ell$ the condition
\begin{equation}
\label{I2}
I_\ell\gg \frac{\hbar\omega_0^3\Gamma^2}{8\pi c^2\Gamma_{ge}^\gamma}
\end{equation}
is obtained.

\subsection{\label{sec3_2}The low-saturation limit}
Two conditions have to be fulfilled for the low-saturation limit. These are (1) a laser intensity sufficiently low for $\rho_{gg}\gg\rho_{ee}$ to hold and (2) that the bandwidth of the laser light is significantly broader than the nuclear transition linewidth $\Gamma_\ell\gg\Gamma$. Due to the second condition, the coherences $\rho_{ge}$ will relax fast to equilibrium, which allows one to set $\dot{\rho}_{ge}=0$. This is known as the adiabatic elimination method \cite{Shore}. By further using that $\rho_{gg}\approx1\gg\rho_{ee}$, one obtains from Eq.~(\ref{twolevelbloch})
\begin{equation}
0=i\frac{\Omega_{eg}}{2}-\rho_{ge}\left(i\Delta\omega+\tilde{\Gamma}\right).
\end{equation}
This transforms to
\begin{equation}
\rho_{ge}=\frac{i\Omega_{eg}/2}{i\Delta\omega+\tilde{\Gamma}}
\end{equation}
and leads, after inserting into the first part of Eq.~(\ref{twolevelbloch}), to the following differential equation for $\rho_{ee}$,
\begin{equation}
\dot{\rho}_{ee}=\frac{\Omega_{eg}^2\tilde{\Gamma}/2}{(\Delta\omega)^2+\tilde{\Gamma}^2}-\rho_{ee}\Gamma,
\end{equation}
which is solved for the boundary condition $\rho_{ee}(t=0)=0$ by
\begin{equation}
\label{lowsat1}
\rho_{ee}=\frac{\Omega_{eg}^2\tilde{\Gamma}/(2\Gamma)}{(\Delta\omega)^2+\tilde{\Gamma}^2}\left(1-e^{-\Gamma t}\right).
\end{equation}
Considering only the case of zero detuning $\Delta\omega=0$, one obtains
\begin{equation}
\label{lowsat3}
\rho_{ee}=\frac{\Omega_{eg}^2}{2\Gamma\tilde{\Gamma}}\left(1-e^{-\Gamma t}\right).
\end{equation}
Note, that Eq.~(\ref{lowsat3}) can also be obtained from Torrey's solution [Eq.~(\ref{Torrey})] by applying the conditions $\tilde{\Gamma}\gg\Gamma\gg\Omega_{eg}$. In this case one has $\lambda\approx(\tilde{\Gamma}-\Gamma)/2$ and $(\Gamma+\tilde{\Gamma})/(2\lambda)\approx 1$. By further using that $\sinh(\lambda t)=(e^{\lambda t}-e^{-\lambda t})/2$ and $\cosh(\lambda t)=(e^{\lambda t}+e^{-\lambda t})/2$ the time dependence is correctly re-obtained.\\[0.2cm]
Further, in the low-saturation limit one has $\Gamma_\ell\gg\Gamma$ and thus, based on Eq.~(\ref{Gammatilde}), $\tilde{\Gamma}\approx\Gamma_\ell/2$. Inserting this into Eq.~(\ref{lowsat3}) and using Eq.~(\ref{Omega01}) for the Rabi frequency $\Omega_{eg}$, the solution of the optical Bloch equations in the low-saturation limit reads:
\begin{equation}
\label{lowsat2}
\colorbox{lightgray}{$
\begin{aligned}
\rho_{ee}=\frac{2\pi c^2 I_\ell \Gamma_{ge}^\gamma}{\hbar \omega_0^3\Gamma\Gamma_\ell}\left(1-e^{-\Gamma t}\right).
\end{aligned}
$}
\end{equation}
The nuclear excitation rate $\Gamma_\text{exc}$ is defined as the number of nuclear excitations per time interval. For the case that all nuclei are in the ground state it can be calculated by $\dot{\rho}_{ee}(0)$.
Based on Eq.~(\ref{lowsat2}) one obtains\footnote{The excitation rate $\Gamma_\text{exc}$ is connected to the absorption cross section $\sigma$ via \cite{Steck} $\Gamma_\text{exc}=\sigma I_\ell/(\hbar\omega_0)$. Therefore one has $\sigma=\lambda_0^2/(2\pi)\cdot \Gamma_{ge}^\gamma/\Gamma_\ell$, with $\lambda_0$ being the wavelength corresponding to the nuclear transition.}
\begin{equation}
\colorbox{lightgray}{$
\begin{aligned}
\Gamma_\text{exc}=\frac{\Omega_{ge}^2}{\Gamma_\ell}=\frac{2\pi c^2 I_\ell \Gamma_{ge}^\gamma}{\hbar \omega_0^3 \Gamma_\ell}.
\end{aligned}
$}
\label{excrate}
\end{equation}

\subsection{Steady-state solution and power broadening}
\noindent In the following, the optical Bloch equations of the two-level system [Eq.~(\ref{twolevelbloch})] are solved for the ``steady-state" case, which denotes the population densities obtained in equilibrium. The steady-state solution of the optical Bloch equations is defined by the conditions $\dot{\rho}_{ee}=\dot{\rho}_{gg}=\dot{\rho}_{ge}=0$, leading to the equations
\begin{equation}
\begin{aligned}
0&=-i\frac{\Omega_{eg}}{2}\left[\rho_{ge}-\rho_{eg}\right]-\rho_{ee}\Gamma,\\
0&=-i\frac{\Omega_{eg}}{2}\left[\rho_{ee}-\rho_{gg}\right]-\rho_{ge}(i\Delta\omega+\tilde{\Gamma}),
\end{aligned}
\end{equation}
which transform to
\begin{equation}
\begin{aligned}
\rho_{ee}&=-i\frac{\Omega_{eg}}{2\Gamma}\left[\rho_{ge}-\rho_{eg}\right],\\
\rho_{ge}&=-i\frac{\Omega_{eg}}{2(i\Delta\omega+\tilde{\Gamma})}\left[\rho_{ee}-\rho_{gg}\right].
\end{aligned}
\end{equation}
By inserting the second equation into the first and using $\rho_{eg}=\rho_{ge}^*$ one obtains
\begin{equation}
\rho_{ee}=\frac{\Omega_{eg}^2\tilde{\Gamma}\left(\rho_{gg}-\rho_{ee}\right)}{2\Gamma\left((\Delta\omega)^2+\tilde{\Gamma}^2\right)}.
\end{equation}
Solving for $\rho_{ee}$ and considering that $\rho_{gg}+\rho_{ee}=1$ results in
\begin{equation}
\label{power}
\rho_{ee}=\frac{\Omega_{eg}^2}{\frac{2\Gamma}{\tilde{\Gamma}}\left((\Delta\omega)^2+\tilde{\Gamma}^2\right)+2\Omega_{eg}^2}.
\end{equation}
Eq.~(\ref{power}) corresponds to a Lorentzian shape of excitation, with the combined width of the laser radiation used for excitation and the width of the nuclear transition. Explicitly, the combined width is obtained as $[4\tilde{\Gamma}^2+4\tilde{\Gamma}\Omega_{eg}^2/\Gamma]^{1/2}$. The effective nuclear transition width can be probed by assuming that the excitation is performed by a laser source with close to zero bandwidth, which allows to take $\Gamma_\ell=0$, thereby leading to $\tilde{\Gamma}=\Gamma/2$. In this case one has
\begin{equation}
\rho_{ee}=\frac{\Omega_{eg}^2}{4(\Delta\omega)^2+\Gamma^2+2\Omega_{eg}^2},
\end{equation}
which corresponds to an effective linewidth (FWHM) of the nuclear transition of $\Gamma_\text{eff}=[\Gamma^2+2\Omega_{eg}^2]^{1/2}$. In case that the Rabi frequency dominates over the nuclear decay rate $\Omega_{eg}\gg\Gamma$, the effective linewidth of the nuclear transition will thus correspond to $\Gamma_\text{eff}\approx\sqrt{2}\Omega_{eg}$. This situation is well known from atomic physics as power broadening \cite{Loudon}.\\[0.2cm]
It is convenient to introduce the saturation intensity $I_\text{sat}$ as the laser intensity that is required in order to obtain $\rho_{ee}=1/4$ in the steady-state case for zero detuning $\Delta\omega=0$ \cite{Steck}. Inserting these conditions into Eq.~(\ref{power}) leads to $\Omega_{eg}^2=\Gamma\tilde{\Gamma}$, which results in combination with Eq.~(\ref{Omega01}) in a value for $I_\text{sat}$ of
\begin{equation}
I_\text{sat}=\frac{\hbar\omega_0^3\Gamma\tilde{\Gamma}}{2\pi c^2\Gamma_{ge}^\gamma}.
\end{equation}
This expression is comparable to the laser intensity required to drive Rabi oscillations for $\Gamma\gg\Gamma_\ell$ given in Eq.~(\ref{I2}), it is, however, different from the required laser intensity for $\Gamma_\ell\gg\Gamma$ [Eq.~(\ref{I1})].


\section{\label{sec3}The optical Bloch equations for multi-level nuclear systems}

\noindent In this section the optical Bloch equations for the case of a nuclear system consisting of multiple nuclear levels as induced due to hyperfine-structure and Zeeman splitting will be discussed. Two different physical scenarios have to be distinguished: (1) nuclei of isolated atoms or ions, e.g., individual laser-cooled ions in a Paul trap and (2) nuclei embedded in a solid-state environment, e.g., in a crystal-lattice structure. These two situations lead to different nuclear level splittings for reasons that will be discussed in the following.\\[0.2cm]
For both, nuclei of isolated atoms or ions and nuclei embedded in a solid-state environment, the nuclear moments (magnetic dipole, electric quadrupole etc.) interact with the fields created by the surrounding electrons, leading to the nuclear HFS. For isolated atoms or ions (1), in addition to the nuclear spin vector $\vec{I}$, also the total angular momentum vector $\vec{J}$ of the electronic shell can be freely oriented, leading to a total of $2\cdot\text{min}(I,J)+1$ hyperfine sub-levels of different energies, where $I$ and $J$ denote the spin and the angular momentum quantum number, respectively \cite{Kopfermann}. Opposed to that, in the solid-state environment (2) the orientation of the electronic shell is determined by the orientation of the lattice. Therefore only the nuclear spin vector can be freely oriented, leading to a splitting into $2I+1$ sub-levels. The details of the splitting will heavily depend on the chemical bonding, which is well known from Mössbauer spectroscopy \cite{Greenwood}.\\[0.2cm]
In addition to the HFS, a Zeeman splitting will occur if an external magnetic field is applied. For the case of free atoms or ions (1), this will lead to an extra splitting of each of the $2\cdot\text{min}(I,J)+1$ hyperfine levels. In the solid-state environment (2) an external magnetic field will induce an additional level shift of each HFS level, but no further level splitting will arise. In the following a quantitative discussion of both physical scenarios will be individually provided.

\subsection{Nuclear splitting for isolated atoms or ions}
\begin{figure}
\begin{center}
\resizebox{0.5\textwidth}{!}{%
  \includegraphics{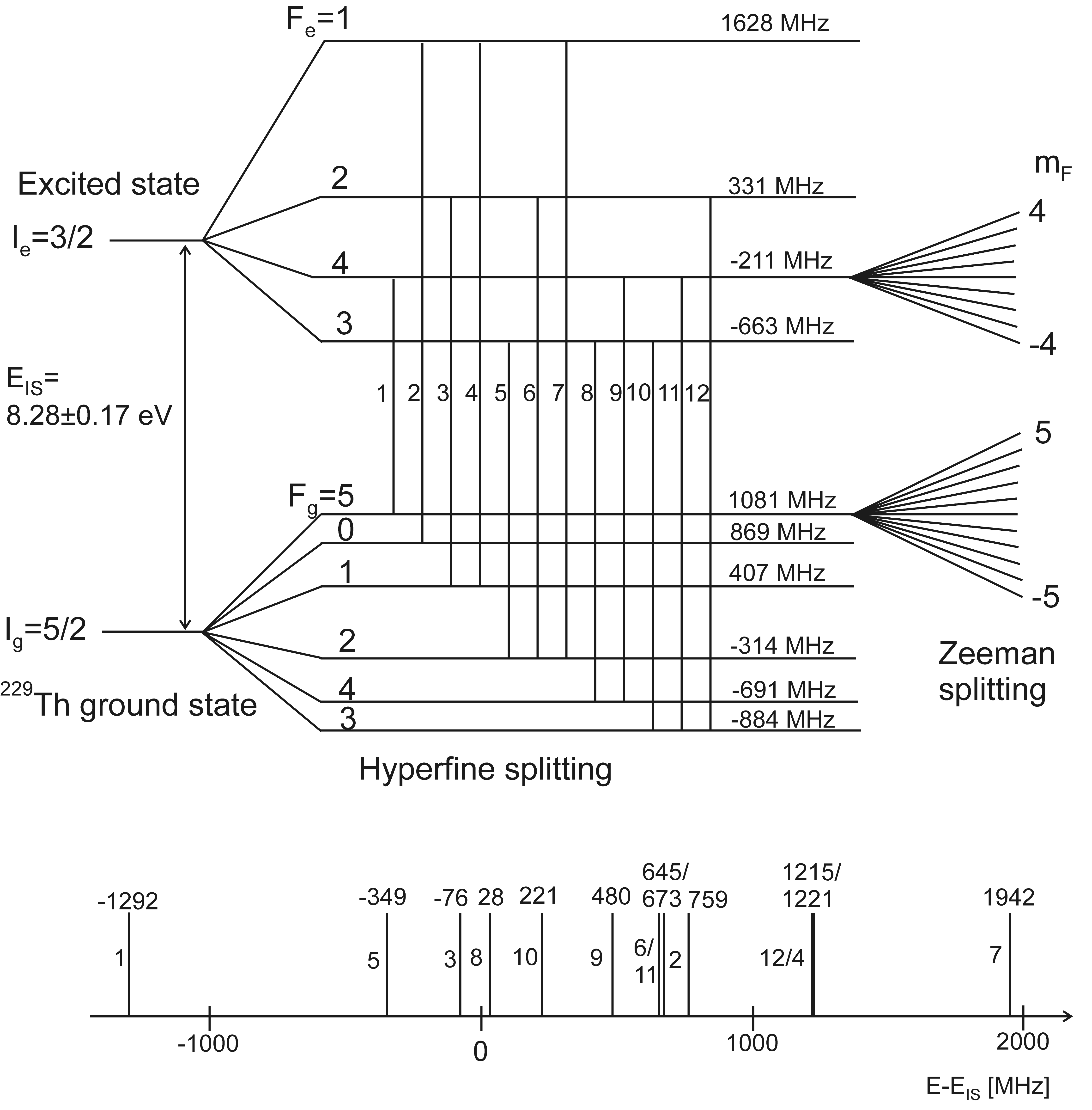}}
\end{center}
  \caption{\footnotesize Hyperfine and Zeeman splitting of nuclear ground and isomeric first excited state of $^{229}$Th$^{3+}$ in the presence of a weak magnetic field (see also Ref.~\cite{Campbell2}). In the lower panel the different hyperfine spectroscopy lines are shown in energetic order with numbers according to their occurrence in the upper panel.}
  \label{splitting1}
\end{figure}

Nuclear levels of isolated atoms or ions, e.g., when laser-cooled and stored in a Paul trap, will experience a HFS due to coupling to the electronic shell. Let $\hat{\vec{I}}$ be the nuclear spin operator of a considered nuclear state, $\hat{\vec{J}}$ be the angular momentum operator of the shell and $\hat{\vec{F}}=\hat{\vec{I}}+\hat{\vec{J}}$ the total angular momentum operator of the system. Further, let $I$, $J$ and $F$ be the corresponding quantum numbers, then the energy shifts $\Delta E_\text{HFS}$ of the different nuclear sub-states (labeled by the quantum number $F$, which takes the $2\cdot\text{min}(I,J)+1$ values from $\vert J-I\vert$ to $J+I$ with steps of $1$) are given by \cite{Kopfermann}
\begin{equation}
\label{HFS}
\Delta E_\text{HFS}=\frac{AC}{2}+\frac{B}{4}\frac{(3/2)C(C+1)-2I(I+1)J(J+1)}{I(2I-1)J(2J-1)}
\end{equation}   
with $C=F(F+1)-J(J+1)-I(I+1)$. Eq.~(\ref{HFS}) contains two energy terms, the first one originating from the magnetic dipole moment and the second one due to the electric quadrupole moment of the nucleus. The parameters $A$ and $B$ are the hyperfine constants, which are related to the nuclear magnetic dipole moment $\mu_I$ as well as the spectroscopic electric quadrupole moment $Q^{(s)}$ via \cite{Kopfermann}
\begin{equation}
\label{hyperfineA}
A=\frac{\mu_IB_\text{el.}(0)}{IJ}
\end{equation}
and (in case of vanishing asymmetry)
\begin{equation}
\label{hyperfineB}
B=Q^{(s)}V_{zz}(0).
\end{equation}
Here $B_\text{el.}(0)$ denotes the magnetic field and $V_{zz}(0)$ the electric field gradient at the site of the nucleus as generated by the electronic shell. Note that $\mu_I$ is defined as the projection of the nuclear magnetic dipole vector $\vec{\mu}_I$ onto the magnetic-field axis for maximum magnetic quantum number $m_I=I$. Therefore $\mu_I=g_I\mu_NI$ holds, with $g_I$ as the nuclear Landé g-factor, which has usually to be experimentally determined, and $\mu_N$ the nuclear magneton. Importantly, Eq.~(\ref{HFS}) is fully symmetric under exchange of $I$ and $J$.\\[0.2cm] 
A further splitting may arise as a consequence of an externally applied magnetic field. In this case, each hyperfine-structure level will split into $2F+1$ sub-states labeled by the magnetic quantum numbers $m_F$. The energy shift experienced by an individual state due to this Zeeman splitting is \cite{Kopfermann}
\begin{equation}
\label{Zeeman1}
\Delta E_\text{Zeeman}=g_F\ \mu_B\ m_F\ B_\text{ext.},
\end{equation}
with $g_F$ the Landé g-factor of the coupled system of shell plus nucleus, $\mu_B$ the Bohr magneton and $B_\text{ext.}$ the magnetic field that is externally applied. $g_F$ can be determined via \cite{Kopfermann}
\begin{equation}
\label{gF}
g_F=g_J\frac{C+2J(J+1)}{2F(F+1)}+g_I\frac{\mu_N}{\mu_B}\frac{C+2I(I+1)}{2F(F+1)},
\end{equation}
with $g_J$ the Landé factor of the electronic shell, given as
\begin{equation}
g_J=1+\frac{J(J+1)-L(L+1)+S(S+1)}{2J(J+1)}\left(g_\text{el.}-1\right).
\end{equation}
Here $L$ denotes the orbital angular momentum of the electronic shell, $S$ the electron spin and $g_\text{el.}\approx2$ the electron g-factor.\\[0.2cm]
In the following, $^{229}$Th$^{3+}$ will be considered as an example. Assuming $^{229}$Th$^{3+}$ to be in its $5F_{5/2}$ electronic ground state, the total angular-momentum quantum number of the shell is $J=5/2$. The quantum numbers of the nuclear spins are $I_g=5/2$ for the ground and $I_e=3/2$ for the first excited state, leading to a HFS into $2I_g+1=6$ levels and $2I_e+1=4$ levels for the ground and excited state, respectively. The hyperfine constants for the $5F_{5/2}$ electronic configuration and the nuclear ground state were experimentally determined to be $A_g/h=82.2$~MHz and $B_g/h=2269$~MHz \cite{Campbell2011}. Taking into account that the electronic shell state remains unchanged, for the nuclear excited state one obtains from Eqs.~(\ref{hyperfineA}) and (\ref{hyperfineB}):
\begin{equation}
\begin{aligned}
\frac{A_e}{h}&=\frac{A_g}{h}\frac{I_g\mu_{Ie}}{I_e\mu_{Ig}}=-141\ \text{MHz}\\
\frac{B_e}{h}&=\frac{B_g}{h}\frac{Q^{(s)}_{e}}{Q^{(s)}_{g}}=1269\ \text{MHz}.
\end{aligned}
\end{equation}
Here $\mu_{I_g}=0.36\mu_N$ and $Q^{(s)}_{g}=3.11$~eb \cite{Safronova2013} as well as $\mu_{I_e}=-0.37\mu_N$ and $Q^{(s)}_{e}=1.74$~eb \cite{Thielking} were used.\\[0.2cm]
It was proposed to use a stretched pair of nuclear hyperfine states in a weak external magnetic field as a clock transition \cite{Campbell2}. Therefore, the nuclear levels carrying the quantum numbers $F_g=5$ for the ground state and $F_e=4$ for the excited state are of particular interest. Based on Eq.~(\ref{gF}) one obtains for the ground and excited state: $g_F\approx3/7$ (independent of $F_g$) and $g_F\approx15/28$ (for $F_e=4$). The corresponding splitting is shown in Fig.~\ref{splitting1}. Taking the photon selection rules of $\Delta F=0,\pm1$ and $\Delta m_F=0,\pm1$ into consideration, the number of individual $\gamma$ lines for the HFS amounts to 12. When, additionally, also an external magnetic field is applied, the Zeeman splitting will lead to a total of 184 individual nuclear transitions.\\[0.2cm]
Similar to the two-level case, discussed in Sec.~\ref{sec2}, the starting point for the derivation of the optical Bloch equations is the quantum optical master equation in Lindblad form, Eq.~(\ref{master}). For the following it is assumed that nuclear ground and excited state consist of sub-states that are energy eigenstates of the unperturbed nuclear Hamiltonian and can be labeled by different quantum numbers. In this case, the density operator $\hat{\rho}(t)$, defined in Eq.~(\ref{densop}), takes the form 
\begin{equation}
\begin{aligned}
\hat{\rho}(t)=&\sum_{e,e'}  \rho_{ee'}\vert e\rangle\langle e'\vert+\sum_{g,e}  \rho_{ge}\vert g\rangle\langle e\vert \\
+&\sum_{g,e}  \rho_{eg}\vert e\rangle\langle g\vert+\sum_{g,g'}  \rho_{gg'}\vert g\rangle\langle g'\vert,
\end{aligned}
\end{equation}
and the nuclear Hamiltonian, defined in Eq.~(\ref{nuclHams1}), can be written as
\begin{equation}
\H_N = \hbar \sum_{e} \omega_{e}|e\rangle\langle e|+
\hbar \sum_{g} \omega_{g}|g\rangle\langle g|.
\end{equation}
Note that different sublevels of the ground and the excited state have different energies due to the hyperfine interaction of the nucleus with the electronic shell, but the resulting differences can be considered to be small in comparison with the nuclear excitation energy.\\[0.2cm]
Like for Eq.~(\ref{hamiltonian3b}) it is assumed that the laser intensity is sufficiently low to set the diagonal terms of the interaction Hamiltonian equal to zero. By further implementing the rotating-wave approximation right from the beginning when dropping the $U$-components, the interaction Hamiltonian $\H_I(t)$, defined in Eq.~(\ref{hamiltonian1}), takes the form
\begin{equation}
\begin{aligned}
&\hat{H}_I(t)= \\
&\hbar\sum_{g,e}\left[V_{ge}|g\rangle\langle e|e^{i\left[\omega_\ell t+\phi(t)\right]}
            +V_{eg}|e\rangle\langle g|e^{-i\left[\omega_\ell t+\phi(t)\right]}\right],\\
\end{aligned}
\label{hamiltonian40}
\end{equation}
where $V_{ge}$ and $V_{eg}$ are defined as previously in Eq.~(\ref{Vdef}).\\[0.2cm]
The Lindblad superoperator $\LL[\hat{\rho}]$, that models the spontaneous decay, takes the explicit form \cite{Scully,Lindblad}
\begin{equation}
\LL[\hat{\rho}]=\sum_{g,e}\Gamma_{ge}\left(\hat{\sigma}_{ge}\hat{\rho}\hat{\sigma}_{ge}^\dagger-\frac{1}{2}\hat{\rho}\hat{\sigma}_{ge}^\dagger\hat{\sigma}_{ge}-\frac{1}{2}\hat{\sigma}_{ge}^\dagger\hat{\sigma}_{ge}\hat{\rho}\right),
\label{lindbladsp20}
\end{equation}
\noindent where $\Gamma_{ge}$ denotes the partial decay rate of a particular excited sub-state to a ground sub-state. As previously $\Gamma_{ge}$ contains radiative as well as non-radiative decay channels, therefore $\Gamma_{ge}=\Gamma_{ge}^\gamma+\Gamma_{ge}^{nr}$. A straightforward calculation, comparable to the one performed in Sec.~\ref{sec2}, leads to the complete set of sub-state optical Bloch equations in the form
\begin{equation}
\begin{aligned}
\dot{\rho}_{ee}=& -i \sum_{g} 
        \frac{\Omega_{eg}}{2}\Big[ \rho_{ge}-\rho_{eg}\Big] -
        \sum_{g} \Gamma_{ge} \rho_{ee}; 
\\
\dot{\rho}_{ge}=&-i\sum_{{e'}} \frac{\Omega_{e'g}}{2} \rho_{e'e} 
          +i\sum_{{g'}} \frac{\Omega_{eg'}}{2} \rho_{gg'}
          -\rho_{ge}(i\Delta\omega_{eg}+\tilde{\Gamma}_{ge});
\\
\dot{\rho}_{gg}=& -i \sum_{e} \frac{\Omega_{eg}}{2}
        \Big[\rho_{eg}-\rho_{ge}\Big] +
        \sum_{e} \Gamma_{ge} \rho_{ee}; 
\\
\dot{\rho}_{ee'}=& -i \sum_{g} \Big[ \frac{\Omega_{eg}}{2} \rho_{ge'}- \frac{\Omega_{e'g}}{2}\rho_{eg} \Big]
-\rho_{ee'}(i \omega_{ee'}+\tilde{\Gamma}_{ee'});
\\
\dot{\rho}_{gg'}=& -i \sum_{e} \Big[ \frac{\Omega_{eg}}{2} \rho_{eg'}- \frac{\Omega_{eg'}}{2} \rho_{ge} \Big] 
-i \omega_{gg'}\rho_{gg'}.
\label{bloch50}
\end{aligned}
\end{equation}

\noindent Here the notation $\omega_{ij}=\omega_i-\omega_j$ as well as $\Delta\omega_{eg}=\omega_\ell-\omega_{eg}$ was used and the decay rates of the coherences $\tilde{\Gamma}_{ge}$ as well as $\tilde{\Gamma}_{ee'}$ were introduced as
\begin{equation}
\begin{aligned}
\tilde{\Gamma}_{ge}&=\sum_{g'}\frac{\Gamma_{g'e}}{2}+\frac{\Gamma_\ell}{2},\\
\tilde{\Gamma}_{ee'}&=\sum_{g} \frac{\Gamma_{ge}+\Gamma_{ge'}}{2}.
\end{aligned}
\end{equation}
For typical laser intensities and linewidths considered for $^{229\text{m}}$Th, the excitation rate per nucleus $\Omega_{eg}^2/\tilde{\Gamma}_{ge}$ will be small compared to the energy differences $\omega_{gg'}$ and $\omega_{ee'}$ between different sub-states of the ground and excited states or in comparison with the decoherence rates $\tilde{\Gamma}_{ee'}$ and $\tilde{\Gamma}_{gg'}$. For example, in Ref.~\cite{Wense3} a laser with intensity of $2\cdot10^5$ W/cm$^2$ and bandwidth $2\pi\cdot 10^{10}$~Hz is considered, leading to an excitation rate of about $10^{-2}$~s$^{-1}$. Opposed to that, a lower intensity of $0.14$ W/cm$^2$ is assumed in Ref.~\cite{Wense2019}, however, at a bandwidth of $2\pi\cdot1$~Hz. At negligible other sources of decoherence this results in an estimated excitation rate of 120~s$^{-1}$ per nucleus. This is significantly smaller than than energy differences between various sublevels of the ground and the excited state (see Fig. 2 as an example) and allows to neglect multi-photon coherences $\rho_{ee'}$ and $\rho_{gg'}$. Therefore the multi-state optical Bloch equations in Eq.~(\ref{bloch50}) transform to \cite{Palffy}
\begin{equation}
\colorbox{lightgray}{$
\begin{aligned}
\dot{\rho}_{ee}&=-i\sum_{g}\frac{\Omega_{eg}}{2}\Big[\rho_{ge} - \rho_{eg} \Big]-\rho_{ee}\sum_g\Gamma_{ge},\\
\dot{\rho}_{ge}&=-i\frac{\Omega_{eg}}{2}\Big[\rho_{ee}-\rho_{gg}\Big]-\rho_{ge}\left( i\Delta\omega_{eg}+\tilde{\Gamma}_{ge}\right),\\
\dot{\rho}_{gg}&=-i\sum_{e}\frac{\Omega_{eg}}{2}\Big[\rho_{eg} - \rho_{ge} \Big]+\sum_{e}\rho_{ee}\Gamma_{ge}.
\label{bloch60}\end{aligned}
$}
\end{equation}
\noindent This set of equations can only be solved numerically. An analytic solution exists, however, for the case of low saturation, which will be discussed in Sec.~\ref{sec4}. Important input parameters are the Rabi frequencies $\Omega_{eg}$ as well as the nuclear sub-level decay rates $\Gamma_{ge}$. The Rabi frequencies are defined as previously in Eq.~(\ref{Omega01}), which leaves us with the derivation of the explicit form of the decay rates $\Gamma_{ge}=\Gamma_{ge}^\gamma+\Gamma_{ge}^{nr}$.\\[0.2cm]
For magnetic multipole radiation, the decay rate $\Gamma_{ge}^\gamma$ between the states $\ket{e}$ and $\ket{g}$ was expressed in terms of the matrix element of the magnetic multipole operator $\langle g \vert \hat{\mathcal{M}}_{L\sigma}\vert e \rangle$ in Eq.~(\ref{magneticdecay}) and, similarly, for electric multipole radiation in Eq.~(\ref{electricdecay}). For the following, rotational symmetry around a quantization axis determined by a constant external field is assumed. This results in the commutation of the total angular momentum operator squared $\hat{\vec{F}}^{_2}$ as well as its projection $\hat{\vec{F}}_z$ onto the quantization axis with the nuclear Hamiltonian $\H_N$. In this case, the energy levels are determined by the quantum numbers $I$, $J$, $F$ and $m_F$. Here $m_F$ is a magnetic quantum number, which labels the $2F+1$ magnetic sublevels. It is therefore convenient to choose the following sublevels of nuclear ground and excited state as basis states:
\begin{equation}
\begin{aligned}
\vert g \rangle&=\vert \alpha_g, J_g, I_g, F_g, m_{F_g}\rangle, \\
\vert e \rangle&=\vert \alpha_e, J_e, I_e, F_e, m_{F_e}\rangle.
\end{aligned}
\end{equation}
All remaining quantum numbers, which describe the nuclear sub-states, were combined in $\alpha$. This allows one to express the matrix element of the magnetic multipole operator $\langle g \vert \hat{\mathcal{M}}_{L\sigma}\vert e \rangle$ in terms of the reduced matrix element by means of the Wigner-Eckart theorem (see, e.g., Eq.~(C.84) of Ref.~\cite{Messiah}):
\begin{equation}
\label{Wignereckart}
\begin{aligned}
&\vert\langle \alpha_g,J_g,I_g,F_g,m_{F_g} \vert \hat{\mathcal{M}}_{L\sigma}\vert \alpha_e,J_e,I_e,F_e,m_{F_e}\rangle\vert\\
=&\left\vert\left(\begin{matrix} F_g & L & F_e \\ -m_{F_g} & \sigma & m_{F_e} \end{matrix}\right)\right\vert \vert\langle \alpha_g,J_g,I_g,F_g \Vert \hat{\mathcal{M}}_{L}\Vert \alpha_e,J_e,I_e,F_e \rangle\vert.
\end{aligned}
\end{equation}
Here the Wigner 3J symbol was introduced in round brackets. Its definition is given, e.g., in Ref.~\cite{Messiah}. By further using Eq.~(C.90) of Ref.~\cite{Messiah} and inserting the resulting expression into Eq.~(\ref{magneticdecay}) one can relate the decay rate $\Gamma_{ge}^\gamma$ to the reduced matrix element of the magnetic multipole operator $\langle \alpha_g, I_g\Vert \hat{\mathcal{M}}_{L} \Vert \alpha_e, I_e\rangle$ as \cite{Messiah}\\
\begin{strip}
\begin{equation}
\label{substatedecay}
\begin{aligned}
\Gamma_{ge}^\gamma&=\frac{2\mu_0}{\hbar}
\frac{(L+1)k^{2L+1}}{L[(2L+1)!!]^2}
\left\vert\langle \alpha_g,J_g,I_g,F_g,m_{F_g} \vert \hat{\mathcal{M}}_{L\sigma}\vert \alpha_e,J_e,I_e,F_e,m_{F_e}\rangle\right\vert^2\\
&=\frac{2\mu_0}{\hbar}
\frac{(L+1)k^{2L+1}}{L[(2L+1)!!]^2}\left\vert\left(\begin{matrix} F_g & L & F_e \\ -m_{F_g} & \sigma & m_{F_e} \end{matrix}\right)\right\vert^2 \left\vert\langle \alpha_g,J_g,I_g,F_g \Vert \hat{\mathcal{M}}_{L}\Vert \alpha_e,J_e,I_e,F_e \rangle\right\vert^2\\
&=\frac{2\mu_0}{\hbar}\frac{(L+1)k^{2L+1}}{L[(2L+1)!!]^2}(2F_e+1)(2F_g+1)\left\vert\biggl\lbrace\begin{matrix} I_g & L & I_e \\ F_e & J & F_g \end{matrix}\biggr\rbrace\right\vert^2 \left\vert \left(\begin{matrix} F_g & L & F_e \\ -m_{F_g} & \sigma & m_{F_e} \end{matrix}\right)\right\vert^2 \left\vert\langle \alpha_g,I_g \Vert \hat{\mathcal{M}}_{L}\Vert \alpha_e,I_e \rangle\right\vert^2.
\end{aligned}
\end{equation}
\end{strip}
\noindent As before, round brackets were used for the Wigner-3J symbol and the Wigner-6J symbol was introduced in curly brackets. The definition of these symbols can be found in Ref.~\cite{Messiah}. For the last equation it was assumed, that the atomic shell state remains unchanged with the angular momentum quantum number $J_g=J_e=J$. The obtained expression can be further simplified by expressing the reduced matrix element of the magnetic multipole operator in terms of the total radiative rate $\Gamma_\gamma$ of the nuclear transition. For this purpose the matrix element is related to the reduced transition probability $B_{\downarrow}(ML)$ via \cite{Ring}
\begin{equation} \label{reducedtransition}
\left\vert\langle \alpha_g,I_g \Vert \hat{\mathcal{M}}_{L}\Vert \alpha_e,I_e \rangle\right\vert^2=(2I_e+1)B_{\downarrow}(ML).
\end{equation}
Inserting Eq.~(\ref{reducedtransition}) into Eq.~(\ref{substatedecay}) and using the definition of the total radiative decay rate for magnetic multipole radiation in terms of the reduced transition probability \cite{Ring}
\begin{equation}\label{totaldecay}
\Gamma_\gamma=\frac{2\mu_0}{\hbar}\frac{(L+1)}{L[(2L+1)!!]^2}k^{2L+1}  B_{\downarrow}(ML)  \;,
\end{equation}
the final expression for the partial radiative decay rates is obtained as
\begin{equation}
\label{radiativedecayrate1}
\colorbox{lightgray}{$
\begin{aligned}
\Gamma_{ge}^\gamma=&(2F_e+1)(2F_g+1)(2I_e+1)\times\\
&\left\vert\biggl\lbrace\begin{matrix} I_g & L & I_e \\ F_e & J & F_g \end{matrix}\biggr\rbrace\right\vert^2 \left\vert \left(\begin{matrix} F_g & L & F_e \\ -m_{F_g} & \sigma & m_{F_e} \end{matrix}\right)\right\vert^2 \Gamma_\gamma.
\end{aligned}$}
\end{equation}
This expression is identical to what is known from atomic physics (see, e.g., Ref.~\cite{King}), with the important difference that $I$ and $J$ appear interchanged and arbitrary multipole order $L$ is considered. A comparable calculation can also be performed for electric multipole radiation, leading to the identical expression. By applying the orthogonality relations of the Wigner symbols (see, e.g., Ref.~\cite{Edmonds}) one can show that the total radiative decay rate is obtained by summation over all ground-level sub-states, just as expected:
\begin{equation}
\Gamma_\gamma=\sum_{F_g, m_{F_g} } \Gamma_{ge}^\gamma=\sum_{F_g, m_{F_g}, \sigma} \Gamma_{ge}^\gamma.
\end{equation}
The sum over $\sigma$ can be added without changing the result, as for defined ground and excited state only one term is non-zero.\\[0.2cm]
Although it might be very hard to calculate the partial non-radiative decay rates $\Gamma_{ge}^{nr}$ explicitly, the sum over all ground states will lead to the total non-radiative decay rate $\Gamma_{nr}$:
\begin{equation}
\Gamma_{nr}=\sum_{F_g,m_{F_g}}\Gamma_{ge}^{nr}.
\end{equation}
The expression of the partial non-radiative decay rates via the total non-radiative decay rate depends on the specific type of the process and cannot be written in the general case. For the demonstration purposes we assume at this point the simplest case that all partial non-radiative decay rates are identical. This results in
\begin{equation}
\label{nonradpartial}
\Gamma_{ge}^{nr}\approx\frac{\Gamma_{nr}}{4JI_g+2(J+I_g)+1}.
\end{equation}
Here, the denominator corresponds to the sum over all possible ground states.\footnote{ With $a=|J-I_g|$ and $b=|J+I_g|$ one has:
\begin{equation}
\begin{aligned}
&\sum_{i=a}^{b}(2i+1)\\
=&(b-a+1)/2\left[(2a+1)+(2b+1)\right]\\
=&b^2-a^2+2b+1\\
=&4JI_g+2(J+I_g)+1.
\end{aligned}
\end{equation}}


\subsection{Nuclear splitting in a solid-state environment}
\begin{figure}
\begin{center}
\resizebox{0.4\textwidth}{!}{%
  \includegraphics{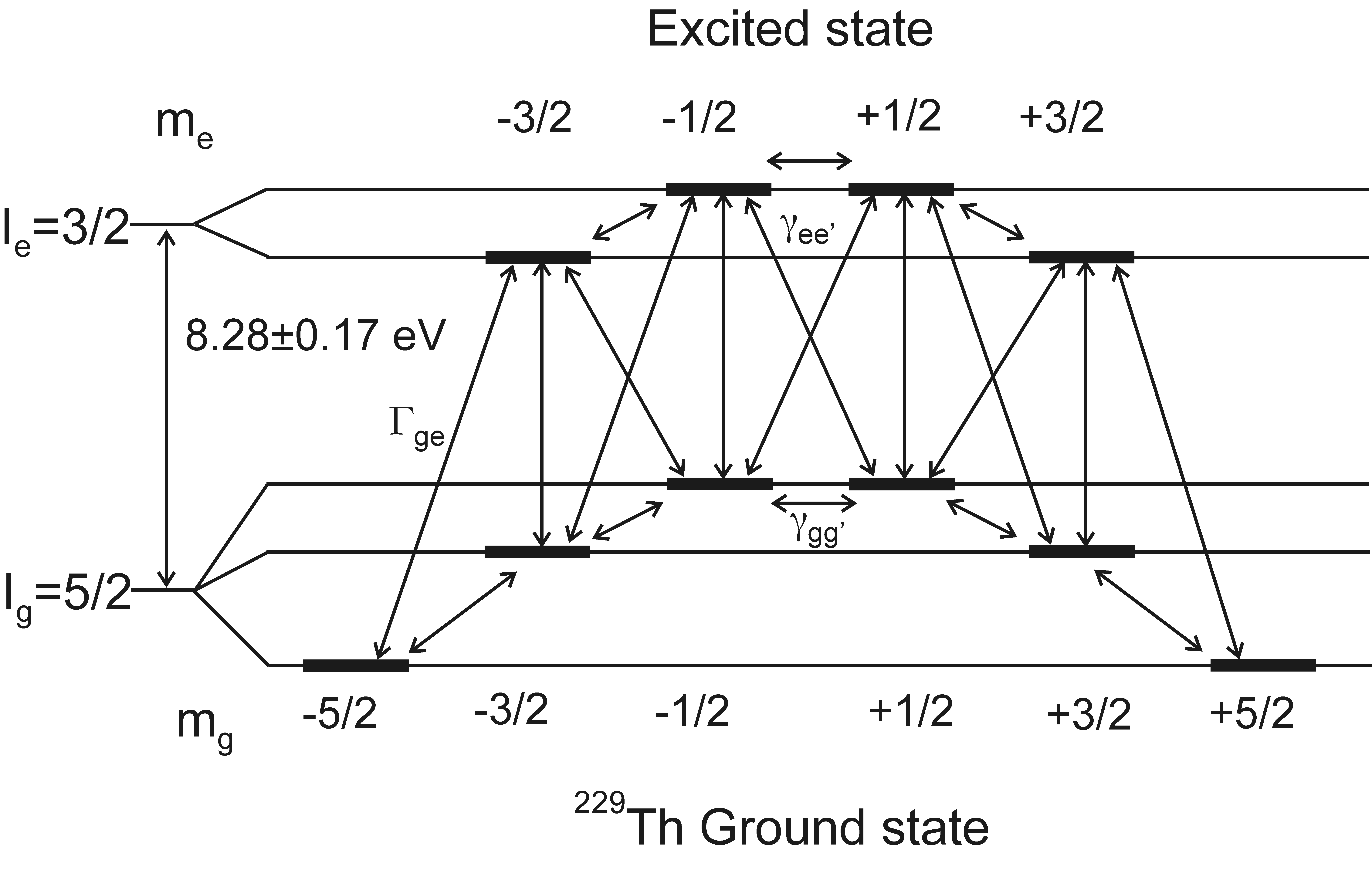}}
\end{center}
  \caption{\footnotesize Example for a nuclear energy splitting as induced by the nuclear quadrupole moment in the presence of an electric field gradient caused by a crystal-lattice environment: the $^{229}$Th nucleus with ground- and first excited state embedded in a CaF$_2$ crystal with charge compensation (see Ref.~\cite{Kazakov1} for details).}
  \label{splitting}
\end{figure}

The laser irradiation of nuclei embedded in a solid-state environment is similar to the case of Mössbauer spectroscopy \cite{Greenwood}. An energy splitting into $2I+1$ sub-states caused by a magnetic field $B$ at the point of the nucleus will arise with energy spacing $\Delta E_\text{Zeeman}=g_I\ \mu_N\ m_I\ B$. This expression equals Eq.~(\ref{Zeeman1}), however with $m_I$ and $g_I$ instead of $m_F$ and $g_F$. A further shift of the magnetic sub-states arises due to the energy of the nuclear electric quadrupole moment in the electric field gradient generated by the electronic shell. This energy shift is quantitatively expressed for vanishing asymmetry by \cite{Greenwood}
\begin{equation}
\label{EQ}
\Delta E_Q=\frac{Q^{(s)} V_{zz}(0)}{4} \frac{3m_I^2-I(I+1)}{I(2I-1)}.
\end{equation}
Here $Q^{(s)}$ denotes the spectroscopic quadrupole moment of the nucleus and $V_{zz}(0)$ the second derivative of the electric potential at the point of the nucleus. The case of non-vanishing asymmetry is discussed, e.g., in Refs.~\cite{Dessovic,Nickerson}.\\[0.2cm]
Of particular interest in this context are $^{229}$Th nuclei embedded in a crystal lattice environment consisting, e.g., of a CaF$_2$ or a LiSrAlF$_6$ crystal \cite{Jeet,Stellmer}. In both cases $^{229}$Th ions will exhibit a $4+$ charge state when grown into the crystal lattice structure. For this reason the electronic shell is closed and the magnetic field at the point of the nucleus is expected to be weak. In case that the electric quadrupole shift, given in Eq.~(\ref{EQ}), dominates the nuclear energy splitting, e.g., when $^{229}$Th$^{4+}$ is embedded in a CaF$_2$ crystal with charge compensation, the splitting will arise as shown in Fig.~\ref{splitting} \cite{Kazakov1}.\\[0.2cm]
The derivation of the optical Bloch equations for the solid-state environment can be found in Appendix B. Here merely the result is given for the case that $\Omega_{eg}^2/\tilde{\Gamma}_{ge}$ is small compared to the energy differences $\omega_{gg'}$ and $\omega_{ee'}$ between different sub-states or in comparison with the decoherence rates $\tilde{\Gamma}_{ee'}$ and $\tilde{\Gamma}_{gg'}$ between sub-states. In this case one obtains
\begin{equation}
\colorbox{lightgray}{$
\begin{aligned}
\dot{\rho}_{ee}=& -i \sum_{g} \frac{\Omega_{eg}}{2}
        \Big[ \rho_{ge}-\rho_{eg}\Big] -
        \sum_{g} \Gamma_{ge} \rho_{ee}\\
         & + \phantom{i} \sum_{e'} \gamma_{ee'} (\rho_{e'e'}-\rho_{ee}); 
\\
\dot{\rho}_{ge}=&-i \frac{\Omega_{eg}}{2} \Big[\rho_{ee}
          - \rho_{gg} \Big]
          -\rho_{ge}(i\Delta\omega_{eg}+\tilde{\Gamma}_{ge});
\\
\dot{\rho}_{gg}=& -i \sum_{e} 
        \frac{\Omega_{eg}}{2}\Big[\rho_{eg}-\rho_{ge}\Big] +
        \sum_{e} \Gamma_{ge} \rho_{ee}\\
        & + \phantom{i} \sum_{g'} \gamma_{gg'} (\rho_{g'g'}-\rho_{gg}). 
\label{bloch51}
\end{aligned}
$}
\end{equation}
Here $\tilde{\Gamma}_{ge}$ is defined as
\begin{equation}
\label{gammatilde10}
\tilde{\Gamma}_{ge}=\sum_{g'}\frac{\Gamma_{g'e}+\gamma_{g'g}}{2}
                    +\sum_{e'}\frac{\gamma_{e'e}}{2}+\frac{\Gamma_\ell}{2}.
\end{equation}
Considering the condition for $\Omega_{eg}^2/\tilde{\Gamma}_{ge}$, the only difference of Eq.~(\ref{bloch51}) to the optical Bloch equations for free atoms or ions, given in Eq.~(\ref{bloch60}), are the additional terms containing the sub-level decay rates $\gamma_{gg'}$ and $\gamma_{ee'}$, which describe the sub-level mixing as induced by the environment. The Rabi frequencies are defined exactly as in Eq.~(\ref{Omega01}), therefore in the following we will focus on the derivation of the explicit form of the decay rates. For magnetic multipole radiation $\Gamma_{ge}^\gamma$ can be expressed in terms of the magnetic multipole operator by means of Eq.~(\ref{magneticdecay}). As previously, rotational symmetry around a quantization axis determined by a constant external field of the environment is assumed. This time, this results in the commutation of the nuclear angular momentum operator squared $\hat{\vec{I}}^{_2}$ as well as its projection $\hat{\vec{I}}_z$ onto the quantization axis with the nuclear Hamiltonian $\H_N$. We choose therefore states with particular quantum numbers $I$ and $m_I$ as basis states. All remaining quantum numbers which describe the nuclear state are combined in a further variable called $\alpha$:
\begin{equation}
\begin{aligned}
\vert g \rangle&=\vert \alpha_g, I_g, m_{I_g}\rangle, \\
\vert e \rangle&=\vert \alpha_e, I_e, m_{I_e}\rangle.
\end{aligned}
\end{equation}
This allows one to express the matrix element of the magnetic multipole operator in terms of the corresponding reduced matrix element by means of the Wigner-Eckart theorem [see also Eq.~(\ref{Wignereckart})]:
\begin{equation}
\label{Wignereckart2}
\begin{aligned}
&\langle \alpha_g,I_g, m_{I_g} \vert \hat{\mathcal{M}}_{L\sigma}\vert \alpha_e,I_e, m_{I_e}\rangle\\
=&(-1)^{I_g-m_{I_g}} \left(\begin{matrix} I_g & L & I_e \\ -m_{I_g} & \sigma & m_{I_e} \end{matrix}\right) \langle \alpha_g,I_g \Vert \hat{\mathcal{M}}_{L}\Vert \alpha_e,I_e \rangle,
\end{aligned}
\end{equation}
where it was assumed that the parity condition for the transition is fulfilled. Inserting this into Eq.~(\ref{magneticdecay}) one obtains
\begin{equation}
\begin{aligned}
\Gamma_{ge}^\gamma=&
\frac{2\mu_0}{\hbar}
\frac{L+1}{L[(2L+1)!!]^2}
k^{2L+1}
\left\vert\left(\begin{matrix} I_g & L & I_e \\ -m_{I_g} & \sigma & m_{I_e} \end{matrix}\right)\right\vert^2\\ &\times\left\vert\langle \alpha_g,I_g \Vert \hat{\mathcal{M}}_{L}\Vert \alpha_e,I_e \rangle\right\vert^2.
\end{aligned}
\end{equation}
Now applying Eqs.~(\ref{reducedtransition}) and (\ref{totaldecay}), the partial radiative decay rate is related to the total radiative decay rate $\Gamma_\gamma$ via
\begin{equation}
\label{gammam}
\colorbox{lightgray}{$
\begin{aligned}
\Gamma_{ge}^\gamma= (2I_e+1)\left\vert \left(\begin{matrix} I_g & L & I_e \\ -m_{I_g} & \sigma & m_{I_e} \end{matrix}\right)\right\vert^2 \Gamma_\gamma.
\end{aligned}
$}
\end{equation}
In this way
\begin{equation}
\Gamma_\gamma=\sum_{m_{I_g}} \Gamma_{ge}^\gamma=\sum_{m_{I_g},\sigma} \Gamma_{ge}^\gamma
\end{equation}
holds. Further, assuming as before that the partial non-radiative decay rates can be approximated as being identical for all sub-states, one has
\begin{equation}
\label{nonradpartial2}
\Gamma_{ge}^{nr}\approx\frac{\Gamma_{nr}}{2I_g+1}.
\end{equation}
The multi-state optical Bloch equations for the solid-state environment [Eq.~(\ref{bloch51})] can, in general, only be solved numerically. However, under certain conditions, which are fulfilled in case of the low-saturation limit (see Sec.~\ref{sec3_2}), an analytic solution exists. This situation will be discussed in the following section. 

\section{\label{sec4}The low-saturation limit considering sub-states}

\noindent In this section the optical Bloch equations of a multilevel system [given in Eq.~(\ref{bloch51})] will be analytically solved under the assumption that the nuclear transition is driven with moderate laser intensities and a laser bandwidth significantly broader than the natural linewidth of the nuclear transition. These conditions are known as the low-saturation limit (see Sec.~\ref{sec3_2}).\\[0.2cm]
In the low-saturation limit the extra terms in Eq.~(\ref{bloch51}), which describe the population mixing of the sub-states, can be dropped. This is the case, as for low laser intensities the ground-level population will practically not change, while for the excited level only the total population is of interest. Therefore Eq.~(\ref{bloch51}) will effectively transform to the optical Bloch equations for isolated atoms or ions, Eq.~(\ref{bloch60}). Further, as the bandwidth $\Gamma_\ell$ of the excitation light is assumed to be significantly larger than the ``bare'' optical decoherence rates caused by the interaction with the environment, from Eq.~(\ref{gammatilde10}) it follows that $\tilde{\Gamma}_{ge}$ is independent of $g$ and $e$ ($\tilde{\Gamma}_{ge}=\tilde{\Gamma}$). Under these conditions, the optical Bloch equations, Eq.~(\ref{bloch60}), read
\begin{align}
\dot{\rho}_{ee}&=-i\sum_{g}\frac{\Omega_{eg}}{2}\Big[\rho_{ge} - \rho_{eg} \Big]-\rho_{ee}\sum_g\Gamma_{ge},
\label{bloch2ee}\\
\dot{\rho}_{ge}&=-i\frac{\Omega_{eg}}{2}\Big[\rho_{ee}-\rho_{gg}\Big]-\rho_{ge}\left( i\Delta\omega_{eg}+\tilde{\Gamma}\right),
\label{bloch2ge}\\
\dot{\rho}_{gg}&=-i\sum_{e}\frac{\Omega_{eg}}{2}\Big[\rho_{eg} - \rho_{ge} \Big]+\sum_{e}\rho_{ee}\Gamma_{ge}.
\label{bloch2gg}\end{align}
For $\Gamma_\ell$ significantly larger than the typical evolution rate of the system, the coherences $\rho_{ge}$ will relax fast to equilibrium. In this case, one can set $\dot{\rho}_{ge}=0$, which is known as the adiabatic elimination method \cite{Shore}. Further using that, in the low-saturation limit, the excited states will be much less populated than the ground states: $\rho_{gg}\gg\rho_{ee}$ and $\sum_{g}\rho_{gg}\approx 1$ and that all sub-states of the ground level will be equally populated [$\rho_{gg}\approx 1/(2I_{g}+1)$], Eq.~(\ref{bloch2ge}) simplifies to
\begin{equation}
0=\frac{i\Omega_{eg}/2}{2I_{g}+1}-\rho_{ge}\left(i\Delta\omega_{eg}+\tilde{\Gamma}\right).
\end{equation}
By solving for $\rho_{ge}$ one obtains
\begin{equation}
\rho_{ge}=\frac{i}{2I_{g}+1}\ \frac{\Omega_{eg}/2 }{i\Delta\omega_{eg}+\tilde{\Gamma}}.
\end{equation}
Inserting this expression into Eq.~(\ref{bloch2ee}) leads to
\begin{equation}
\dot{\rho}_{ee}=\frac{1}{2(2I_{g}+1)}\sum_{g}\frac{\tilde{\Gamma}\Omega_{eg}^{2}}{(\Delta\omega_{eg})^{2}+\tilde{\Gamma}^{2}}-\rho_{ee}\sum_g\Gamma_{ge}.
\label{diffee}
\end{equation}
From previous discussions it is known that
\begin{equation}
\sum_{g}\Gamma_{ge}=\sum_g\Gamma_{ge}^\gamma+\sum_g\Gamma_{ge}^{nr}=\Gamma_\gamma+\Gamma_{nr}=\Gamma.
\label{Gammatot}
\end{equation}
The differential equation Eq.~(\ref{diffee}) is then solved for $\rho_{ee}(t=0)=0$ by
\begin{equation}
\label{rhoeeme}
\rho_{ee}=\frac{1-e^{-\Gamma t}}{2\Gamma(2I_{g}+1)}\sum_{g}\frac{\tilde{\Gamma} \Omega_{eg}^{2}}{(\Delta\omega_{eg})^{2}+\tilde{\Gamma}^{2}}.
\end{equation}
In the following, the energy differences between different sub-states of the excited and ground states ($\omega_{ee'}$ and $\omega_{gg'}$) are assumed to be small in comparison either with the total decoherence rate $\tilde{\Gamma}$, or with the detuning $\Delta \omega_{eg}$. This allows to replace $\omega_{eg}$ by some characteristic averaged value $\omega_0$ and $\Delta\omega_{eg}$ by $\Delta\omega$. Thus the factor ${\tilde{\Gamma}}/{(\Delta\omega^{2}+\tilde{\Gamma}^{2})}$ can be taken out of the sum of Eq.~(\ref{rhoeeme}). Consequently, from Eq.~(\ref{rhoeeme}) the total excited state population density $\rho_\text{exc}$ is obtained by summation over all sub-states $e$ and reads
\begin{equation}
\begin{aligned}
\rho_\text{exc}&=\sum_{e}\rho_{ee}\\
&=\frac{\tilde{\Gamma}\left[1-e^{-\Gamma t}\right]}{2\Gamma(2I_{g}+1)\left((\Delta\omega)^{2}+\tilde{\Gamma}^{2}\right)}\sum_{g,e}\Omega_{eg}^{2}.
\end{aligned}
\label{rhoexc}
\end{equation}
Using the definition of the Rabi frequencies $\Omega_{eg}$ as given in Eq.~(\ref{Omega01}), the sum over the Rabi frequencies squared is explicitly obtained as
\begin{equation}
\label{sumeg}
\begin{aligned}
\sum_{g,e} \Omega_{eg}^{2}=&\frac{2\pi c^2I_\ell}{\hbar\omega_0^3}\sum_{g,e}\Gamma_{ge}^\gamma\\
=&(2I_e+1)\frac{2\pi c^2I_\ell\Gamma_\gamma}{\hbar\omega_0^3}.
\end{aligned}
\end{equation}
Here $\Gamma_{ge}^\gamma$ as defined in Eq.~(\ref{gammam}) was used. By inserting Eq.~(\ref{sumeg}) to express the sum over the Rabi frequencies in Eq.~(\ref{rhoexc}), the new expression for $\rho_\text{exc}$ is
\begin{equation}
\label{rhotot2}
\rho_\text{exc}=\frac{\pi c^2 I_\ell\Gamma_\gamma}{\hbar \omega_0^3\Gamma}\frac{2I_e+1}{2I_g+1}\frac{\tilde{\Gamma}\left(1-e^{-\Gamma t}\right)}{(\Delta\omega)^2+\tilde{\Gamma}^2}.
\end{equation}
Considering the case of zero detuning of the laser light with respect to the nuclear transition $\Delta\omega=0$ and using that, based on Eq.~(\ref{Gammatilde}), for the low-saturation limit with $\Gamma_\ell\gg\Gamma$ one has $\tilde{\Gamma}\approx\Gamma_\ell/2$, one arrives after multiplication with the total number of irradiated nuclei $N_0$ at the final expression for the time dependent number of nuclear excitations as
\begin{equation}
\label{Nt}
\colorbox{lightgray}{$
\begin{aligned}
N_\text{exc}(t)=\frac{2\pi c^2 I_\ell \Gamma_\gamma N_0}{\hbar \omega_0^3\Gamma\Gamma_\ell}\frac{2I_e+1}{2I_g+1}\left(1-e^{-\Gamma t}\right).
\end{aligned}
$}
\end{equation}
This expression is comparable to Eq.~(\ref{lowsat2}) obtained for the two-level system after multiplication with $(2I_e+1)/(2I_g+1)$. It is emphasized that the same result can also be obtained via the Einstein rate equations, which are valid for the considered case of the low-saturation limit and a transition linewidth smaller than the width of the laser light (see for example Ref.~\cite{Wense3}\footnote{For a Lorentzian spectral shape, the laser intensity $I_\ell$ is related to the spectral energy density $\rho^\omega$ of the laser light at resonance via $I_\ell=\frac{\pi c}{2}\rho^\omega\Gamma_\ell$.}).\\[0.2cm]
As previously, the nuclear excitation rate $\Gamma_\text{exc}$ can be estimated based on Eq.~(\ref{Nt}) as $\dot{N}_\text{exc}(0)$ to be
\begin{equation}
\colorbox{lightgray}{$
\begin{aligned}
\Gamma_\text{exc}=\frac{2\pi c^2 I_\ell \Gamma_\gamma N_0}{\hbar \omega_0^3 \Gamma_\ell}\frac{2I_e+1}{2I_g+1},
\end{aligned}
$}
\label{excrate2}
\end{equation}
which is in full correspondence with Eq.~(\ref{excrate}) obtained for the two-level system.
\section{Nuclear two-photon excitation \label{sec7}}
\noindent The theory of nuclear two-photon excitation was extensively discussed by Romanenko \textit{et al.} in 2012 \cite{Romanenko}. The concept is intriguing, as it would allow to use the easier accessible wavelength of about 300~nm for laser excitation. Also, if a frequency comb is used for spectroscopy, all comb-modes would contribute in a two-photon excitation scheme. In the following, two-photon excitation for pure nuclear laser interaction will be considered, corresponding to the case of a bare nucleus without taking shell-related effects into account. In this case, the interaction Hamiltonian contains the nuclear current density operator only, just as defined in Eq.~(\ref{hamiltonian1}).\\[0.2cm]
For simplicity it is assumed that the system is a two-level system, as discussed in Sec.~\ref{sec2}. In case of two-photon excitation the diagonal terms of $\hat{H}_I$ have to be taken into consideration. Thus Eq.~(\ref{hamiltonian3b}) becomes
\begin{equation}
\label{hamiltonian3c}
\begin{aligned}
\hat{H}_I^{0}&= \hbar \left(V_{g}|g\rangle\langle g|+V_{e}|e\rangle\langle e|+U_{ge}|g\rangle\langle e|+ V_{eg}|e\rangle\langle g|\right), \\
\hat{H}_I^{0\dagger}&= \hbar \left(V^*_{g}|g\rangle\langle g|+V^*_{e}|e\rangle \langle e|+U^*_{ge}|e\rangle \langle g|+ V^*_{eg}|g\rangle \langle e|\right).\\
\end{aligned}
\end{equation}
Here $V_{g}$ and $V_{e}$ are defined as
\begin{equation}
\label{Vdef2}
V_{g}=\frac{\langle g\vert \hat{H}_I^{0}\vert g\rangle}{\hbar},\quad V_{e}=\frac{\langle e\vert \hat{H}_I^{0}\vert e\rangle}{\hbar}.\\
\end{equation}
In Ref.~\cite{Romanenko} it was shown that, after transformation into a rotating frame, similar to the transformation applied in Eq.~(\ref{rotframe}), but with a frequency of $2\omega_\ell$ for the two-photon excitation, keeping terms proportional to $V_{eg}/\omega_\ell$ and averaging over fast oscillation, the optical Bloch equations for two photon excitation effectively read\footnote{The expansion coefficients used in Ref.~\cite{Romanenko} can be transformed to the density matrix formalism using the definitions $\rho_{ee}=c_ec_e^*$, $\rho_{gg}=c_gc_g^*$, as well as $\rho_{ge}=c_gc_e^*$ and $\rho_{eg}=c_ec_g^*$.}
\begin{equation}
\begin{aligned}
\dot{\rho}_{ee}&=-\dot{\rho}_{gg}=-i\Big[\tilde{V}_{eg}\rho_{ge} - \tilde{V}_{ge}\rho_{eg}\Big]-\rho_{ee}\Gamma,\\
\dot{\rho}_{ge}&=\dot{\rho}_{eg}^*=-i\tilde{V}_{ge}\Big[\rho_{ee}-\rho_{gg}\Big]-\rho_{ge}\Big[i\left(\delta+S_{eg}\right)+\tilde{\Gamma}\Big].
\end{aligned}
\label{bloch6}
\end{equation}
Here $\delta$ is defined as the two-photon detuning $\delta=\omega_0-2\omega_\ell$, with $\omega_\ell$ the angular frequency of the laser light, which is assumed to be close to half of the transition frequency $\omega_0$. Considering two-photon excitation of nuclei at rest, using a propagating monochromatic wave, $\tilde{V}_{eg}$ and $S_{eg}$ take the following form \cite{Romanenko}:
\begin{equation}
\begin{aligned}
\tilde{V}_{eg}&=\frac{1}{\omega_\ell}V_{eg}\left(V_e-V_g\right)\\
S_{eg}&=\frac{8}{3}\frac{\vert V_{eg}\vert^2}{\omega_\ell}.
\end{aligned}
\end{equation}
In analogy to Eq.~(\ref{twolevelbloch}), Eq.~(\ref{bloch6}) can be transformed to
\begin{equation}
\label{twolevelbloch2}
\colorbox{lightgray}{$
\begin{aligned}
\dot{\rho}_{ee}&=-\dot{\rho}_{gg}=-i\frac{\tilde{\Omega}_{eg}}{2}\Big[\rho_{ge}-\rho_{eg}\Big]-\rho_{ee}\Gamma,\\
\dot{\rho}_{ge}&=\dot{\rho}_{eg}^*=-i\frac{\tilde{\Omega}_{eg}}{2}\Big[\rho_{ee}-\rho_{gg}\Big]-\rho_{ge}\Big[i\left(\delta+S_{eg}\right)+\tilde{\Gamma}\Big],
\end{aligned}
$}
\end{equation}
with the two-photon Rabi frequency defined as $\tilde{\Omega}_{eg}=2\vert\tilde{V}_{eg}\vert$. In the following, the explicit form of $\tilde{\Omega}_{eg}$ and $S_{eg}$ will be derived.\\[0.2cm]
Based on Eq.~(\ref{Vdef}) in combination with Eq.~(\ref{nonreduced1}) one obtains
\begin{equation}
\vert V_{eg}\vert=\sqrt{\frac{\pi c^2 I_\ell \Gamma_{ge}^\gamma}{2\hbar\omega_0^3}}.
\end{equation}
In order to determine $V_g$ and $V_e$ based on their definitions, Eq.~(\ref{Vdef2}), $\langle g\vert \hat{H}_I^{0}\vert g\rangle$ and $\langle e\vert \hat{H}_I^{0}\vert e\rangle$ has to be calculated. Considering first the matrix element for an individual magnetic sub-state with identical initial and final state, one obtains for the magnetic dipole term ($L=1$) from Eq.~(\ref{RabiFreq1})
\begin{equation}
\begin{aligned}
&\vert\langle\alpha,I,m_I\vert \hat{H}_{I\ M1}^{0\ 0} \vert\alpha,I,m_I\rangle\vert\\
=&\sqrt{\frac{4\pi}{3}}B_{0} \vert\langle\alpha,I,m_I\vert \hat{\mathcal{M}}_{10} \vert\alpha,I,m_I\rangle\vert.
\end{aligned}
\end{equation}
By further applying the Wigner-Eckart theorem [Eq.~(\ref{Wignereckart2})] as
\begin{equation}
\begin{aligned}
&\vert\langle\alpha,I,m_I\vert \hat{\mathcal{M}}_{10}\vert\alpha,I,m_I\rangle\vert\\
=&\left\vert\left(\begin{matrix} I & 1 & I \\ -m_{I} & 0 & m_{I} \end{matrix}\right)\right\vert \vert\langle\alpha,I\vert\vert \hat{\mathcal{M}}_{1}  \vert\vert\alpha,I\rangle\vert,
\end{aligned}
\end{equation}
one obtains
\begin{equation}
\label{Hjm}
\begin{aligned}
&\vert\langle\alpha,I,m_I\vert \hat{H}_{I\ M1}^{0\ 0}\vert\alpha,I,m_I\rangle\vert\\
=&\sqrt{\frac{4\pi}{3}}B_{0}\left\vert\left(\begin{matrix} I & 1 & I \\ -m_{I} & 0 & m_{I} \end{matrix}\right)\right\vert \vert\langle\alpha,I\vert\vert \hat{\mathcal{M}}_{1} \vert\vert\alpha,I\rangle\vert.
\end{aligned}
\end{equation}
Here the interaction matrix element of the $M1$-interaction Hamiltonian was expressed in terms of the magnetic dipole operator $\hat{\mathcal{M}}_{1}$. As the dominant contribution of the energy originates from the lowest multipole order, only the $M1$ case is taken into consideration. Making again use of the Wigner-Eckart theorem, Eq.~(\ref{Wignereckart2}), $\vert\langle\alpha,I\vert\vert \hat{\mathcal{M}}_{1} \vert\vert\alpha,I\rangle\vert$ can be expressed in terms of the magnetic dipole moment $\mu$ of the nuclear state as follows \cite{Ring}:
\begin{equation}
\label{bjm1}
\begin{aligned}
&\vert\langle\alpha,I\vert\vert \hat{\mathcal{M}}_{1} \vert\vert\alpha,I\rangle\vert\\
=&\left\vert\left(\begin{matrix} I & 1 & I \\ -I & 0 & I \end{matrix}\right) \right\vert^{-1}\vert\langle\alpha,I,I\vert \hat{\mathcal{M}}_{1} \vert\alpha,I,I\rangle\vert\\
=&\sqrt{\frac{3}{4\pi}}\left\vert\left(\begin{matrix} I & 1 & I \\ -I & 0 & I \end{matrix}\right) \right\vert^{-1}\vert\mu\vert,
\end{aligned}
\end{equation}
where in the last step the definition of the magnetic dipole moment $\mu$ was used as \cite{Ring}
\begin{equation}
\mu=\sqrt{\frac{4\pi}{3}}\langle\alpha,I,I\vert \hat{\mathcal{M}}_{1} \vert\alpha,I,I\rangle.
\end{equation}
By inserting Eq.~(\ref{bjm1}) into Eq.~(\ref{Hjm}) and using the explicit form of the Wigner-3J symbol:
\begin{equation}
\left\vert\left(\begin{matrix} I & 1 & I \\ -m_{I} & 0 & m_{I} \end{matrix}\right)\right\vert=\frac{\vert m_I\vert }{I(I+1)},
\end{equation}
one obtains
\begin{equation}
\vert\langle\alpha,I,m_I\vert \hat{H}_{I\ M1}^{0\ 0}\vert\alpha,I,m_I\rangle\vert=B_0\frac{\vert m_I\vert }{I}\vert\mu\vert.
\end{equation}
This leads to 
\begin{equation}
\begin{aligned}
\langle g\vert \hat{H}_{I\ 10}^0\vert g\rangle=&B_0\frac{ m_{I_g}}{I_g} \mu_g,\\
\langle e\vert \hat{H}_{I\ 10}^0\vert e\rangle=&B_0\frac{ m_{I_e}}{I_e} \mu_e,
\end{aligned}
\end{equation}
with $\mu_g$ and $\mu_e$ the magnetic dipole moments of the ground and excited nuclear state, respectively.\\[0.2cm]
Based on the above considerations, a straightforward calculation reveals that the two-photon Rabi frequency $\tilde{\Omega}_{eg}$ as well as $S_{eg}$ take the following form close to the two-photon resonance $\omega_\ell\approx\omega_0/2$:
\begin{equation}
\label{Omegatwophoton}
\colorbox{lightgray}{$
\begin{aligned}
\tilde{\Omega}_{eg}&=\sqrt{\frac{4\pi I_\ell^2 \Gamma_{ge}^\gamma}{\hbar^3\epsilon_0c\omega_0^5}}\left\vert \frac{ m_{I_e}}{I_e}\mu_e-\frac{ m_{I_g}}{I_g} \mu_g\right\vert\\
S_{eg}&=\frac{8\pi}{3}\frac{I_\ell c^2\Gamma_{ge}^\gamma}{\hbar\omega_0^4}.\\
\end{aligned}
$}
\end{equation}
Here Eq.~(\ref{Itot1}) for the laser intensity was used. Note, that the two-photon Rabi frequency is proportional to the laser intensity, while the single-photon Rabi frequency, Eq.~(\ref{Omega01}), only depends on the square-root of the laser intensity.\\[0.2cm]
The two-photon optical Bloch equations for a two-level system without magnetic sub-states, Eq.~(\ref{twolevelbloch2}), possess a complete analytical solution, which is obtained by substituting the laser detuning with $\delta+S_{eg}$ in Ref.~\cite{Noh}. For simplicity, only the low-saturation limit is discussed, following the same procedure already presented in Sec.~\ref{sec3_2}. Applying, as before, the adiabatic elimination method \cite{Shore}, which allows to set $\dot{\rho}_{ge}$ equal to zero and using that $\rho_{gg}\approx1\gg\rho_{ee}$, one obtains as a differential equation for $\rho_{ee}$:
\begin{equation}
\dot{\rho}_{ee}=\frac{\tilde{\Omega}_{eg}^2\tilde{\Gamma}/2}{\left(\delta+S_{eg}\right)^2+\tilde{\Gamma}^2}-\rho_{ee}\Gamma,
\end{equation}
which is solved for the boundary condition $\rho_{ee}(t=0)=0$ by
\begin{equation}
\colorbox{lightgray}{$
\begin{aligned}
\rho_{ee}=\frac{\tilde{\Omega}_{eg}^2\tilde{\Gamma}/(2\Gamma)}{(\delta+S_{eg})^2+\tilde{\Gamma}^2}\left(1-e^{-\Gamma t}\right).
\end{aligned}
$}
\end{equation}
For the low-saturation limit one has $\Gamma_\ell\gg\Gamma$ and thus, based on Eq.~(\ref{Gammatilde}), $\tilde{\Gamma}\approx\Gamma_\ell/2$.\\[0.2cm]
The excitation rate $\Gamma_\text{exc}$ per nucleus can be approximated as $\dot{\rho}_{ee}(0)$ and, for the case of zero two-photon detuning, ($\delta=0$) one obtains
\begin{equation}
\Gamma_\text{exc}=\frac{\tilde{\Omega}_{eg}^2\tilde{\Gamma}}{2(S_{eg}^2+\tilde{\Gamma}^2)}\approx\frac{\tilde{\Omega}_{eg}^2}{\Gamma_\ell}.
\end{equation}
In the last approximation it was used that, for realistic parameters, $\tilde{\Gamma}\gg S_{eg}$ holds. By inserting expression (\ref{Omegatwophoton}) for $\tilde{\Omega}_{eg}$, the nuclear two-photon excitation rate can be estimated as
\begin{equation}
\colorbox{lightgray}{$
\begin{aligned}
\Gamma_\text{exc}\approx\frac{4\pi I_\ell^2\Gamma_{ge}^\gamma}{\hbar^3\epsilon_0c\omega_0^5\Gamma_\ell}  \left( \frac{ m_{I_e}}{I_e}\mu_e-\frac{ m_{I_g}}{I_g}\mu_g\right)^2.
\end{aligned}
$}
\label{twophotonrate}
\end{equation}
Similar to the single-photon cross section \cite{Steck}, the two-photon cross section $\sigma$ is connected to the excitation rate via $\Gamma_\text{exc}=\sigma I_\ell^2/(\hbar\omega_0)^2$. Thus one obtains
\begin{equation}
\sigma=\frac{\lambda_0^3}{2\pi^2}\frac{1}{\hbar\epsilon_0 c^4}\frac{\Gamma_{ge}^\gamma}{\Gamma_\ell}\left( \frac{ m_{I_e}}{I_e}\mu_e-\frac{ m_{I_g}}{I_g}\mu_g\right)^2.
\end{equation} 
The probability for direct nuclear two-photon excitation is generally extremely low. The reason is that for nuclei, the magnetic dipole moments $\mu_g$ and $\mu_e$ are of the order of the nuclear magneton $\mu_N$, while for atomic shell transitions the Bohr magneton $\mu_B$ has to be inserted instead. As the ratio of the nuclear magneton to the Bohr magneton is $\mu_N/\mu_B\approx m_e/m_p\approx5.4\cdot10^{-4}$ and the excitation rate scales with the magnetic moments squared, the probability for direct nuclear two-photon excitation has to be expected to be about 7 orders of magnitude suppressed compared to two-photon excitation of atomic shell states. As from the excitation of atomic shell states it is known, that two-photon excitation is usually about four orders of magnitude less efficient than one-photon excitation, there is a total of 11 orders of magnitude that have to be bridged between one-photon and two-photon excitation of the nucleus. This makes the direct two-photon excitation of nuclei very inefficient, even when considering the advantage of a longer wavelength that could be used for excitation. However, it has been argued (e.g., in Refs.~\cite{Romanenko,Bilous2018,Porsev2}) that the excitation rate might be drastically enhanced due to shell-related effects that have not been taken into consideration in the above discussion.\\[0.2cm] 

\section{\label{sec8}The case of $^{229\text{m}}$Th}
\noindent In the previous sections the general theory of nuclear laser excitation was introduced. In the following, the special case of $^{229\text{m}}$Th will be discussed, which involves the lowest known nuclear excitation energy. The isomer (spin and parity 3/2$^+$, Nilsson quantum numbers [631]) to ground state (5/2$^+$ [633]) transition is of multipolarity $M1+E2$ with a recently measured energy of $(8.28\pm0.17)$~eV \cite{Seiferle4}. For $\gamma$ decay and IC in neutral thorium, considered in this work, the $E2$ channel is negligible \cite{Bilous2} and only the $M1$ multipolarity will be taken into consideration. From the theory of multipole radiation [Eq.~(\ref{totaldecay})], the radiative transition rate $\Gamma_\gamma$ for a magnetic transition of multipolarity $L=1$ is obtained to be
\begin{equation}
\Gamma_\gamma=\frac{4}{9}\frac{\mu_{0}k^3}{\hbar}B_{\downarrow}(M1).
\label{A}
\end{equation}
In nuclear physics it is common to measure the reduced transition probability $B_{\downarrow}(ML)$ in Weisskopf units. One Weisskopf unit (1~W.u.) corresponds to the value that $B_{\downarrow}(ML)$ would take when approximated based on the nuclear single-particle shell model. For arbitrary magnetic multipole order $L$ the Weisskopf unit is defined as \cite{Weisskopf}
\begin{equation}
1\ \mathrm{W.u.}(ML)=\frac{10}{\pi}\left(\frac{3}{3+L}\right)^2R^{2L-2}\mu_{N}^{2}.
\end{equation}
Here $R=1.2 A_N^{1/3}$~fm is the nuclear radius (where $A_N$ denotes the mass number) and $\mu_{N}=5.051\cdot10^{-27}$~J/T denotes the nuclear magneton. For the considered case of $L=1$ one obtains $1\ \mathrm{W.u.}=1.79$~$\mu_{N}^{2}$. There is no complete agreement on theoretical predictions of $B_{\downarrow}(M1)$ for $^{229\mathrm{m}}$Th in literature. The predicted values differ by a factor of about 10 between $\sim5.0\cdot10^{-3}$~W.u. and $4.55\cdot10^{-2}$~W.u. \cite{Tkalya4,Ruchowska,Minkov,Minkov2}. Here we conservatively assume $B_{\downarrow}(M1)=5\cdot10^{-3}$~W.u., which corresponds to the most recent value \cite{Minkov2}. Further, it is on the lower limit of the expected reduced transition probability and for this reason leads to the smallest photonic coupling. With this assumption one obtains from Eq.~(\ref{A}) for the $^{229}$Th isomer- to ground state transition $\Gamma_\gamma\approx 9.0\cdot10^{-5}$~Hz, corresponding to a radiative lifetime of $\tau_\gamma=1/\Gamma_\gamma\approx 1.1\cdot10^4$~s.\\[0.2cm]
Such a long lifetime would occur if only the radiative decay channel were present. However, for $^{229\mathrm{m}}$Th there is a strong coupling between the nucleus and the electronic shell, which leads to a rapid isomeric decay via IC, if energetically allowed \cite{Strizhov,Karpeshin1,Tkalya4}. The isomeric energy of $\sim8.3$~eV is above the first ionization potential of thorium of 6.3~eV \cite{Koehler}. Therefore isomeric decay via IC is energetically allowed in the neutral thorium atom, and was experimentally confirmed \cite{Wense1,Seiferle3}. Only if thorium is charged, the IC decay channel is suppressed. However, even in this case a non-radiative decay via, e.g., electronic bridge mechanisms, might be dominant \cite{Strizhov}. The IC coefficient $\alpha_{ic}$ for neutral thorium was numerically calculated, leading to $\alpha_{ic}\approx 10^9$ \cite{Strizhov,Karpeshin1,Tkalya4}. The result is a shortened isomeric lifetime under IC of $\tau_\mathrm{ic}=\tau_\gamma/\alpha_{ic}\approx 10\ \mu\mathrm{s}$, as was recently experimentally verified \cite{Seiferle3}. Nuclear laser excitation of neutral atoms of $^{229}$Th was proposed in Refs.~\cite{Wense3,Wense4,Wense2019}. The concept makes use of the short IC decay channel, which allows to trigger the IC electron detection in coincidence with the laser pulses, thereby leading to a high signal-to-background ratio as well as short scanning times and offering the potential for a significantly improved $^{229\text{m}}$Th energy determination.\\[0.2cm]
In the following, the case of nuclear laser excitation of $^{229}$Th ions in a Paul trap is considered as proposed for the nuclear clock concept \cite{Peik1,Campbell2}. For this case it is assumed that the IC decay channel is suppressed ($\alpha_{ic}=0$) and that no significant electronic bridge channels occur, leading to a total lifetime that equals the radiative lifetime: $\Gamma=\Gamma_\gamma\approx9.0\cdot10^{-5}$~Hz. Further, in such a scenario the bandwidth of the laser light is expected to be significantly broader than the nuclear transition linewidth $\Gamma_\ell\gg\Gamma$ and for the following calculations the nuclear transition is assumed to be excited by one comb tooth of a VUV frequency comb (generated via HHG) providing $\Gamma_\ell=2\pi\cdot1$~Hz bandwidth. For the operation of a single-ion nuclear clock, it is important to drive Rabi oscillations of the nuclear system as it allows to use the Ramsey interrogation scheme, thereby improving the expected clock stability to a value limited by the quantum-projection noise \cite{Ludlow}. For the given parameters, the laser intensity required to drive Rabi oscillations is estimated based on Eq.~(\ref{I1}) to be
\begin{equation}
\label{Thestimate}
I_\ell\gg \frac{\hbar\omega_0^3\Gamma_\ell^2}{8\pi c^2\Gamma_\gamma}\approx 4.0\cdot10^{-3}\ \text{W}/\text{cm}^2.
\end{equation}
The identical condition also follows from Eq.~(\ref{excrate}), if $\Gamma_\text{exc}=\Gamma_\ell=2\pi\cdot1$~Hz is assumed. A laser intensity of significantly larger than 4~mW/cm$^2$ in an individual comb tooth of a frequency comb with $2\pi\cdot1$~Hz bandwidth at about 150~nm is experimentally conceivable. For comparison: the maximum reported outcoupled power of an individual comb tooth in the XUV (at about 97~nm) amounts to $\sim5$~nW \cite{Zhang}. This was achieved in the 11th harmonic of an Yb-doped fiber frequency comb, generated via cavity-enhanced HHG. For nuclear laser excitation of $^{229}$Th, the 7th harmonic (around 150~nm) would be required. Assuming the same outcoupled power for the 7th harmonic like for the 11th harmonic, a laser intensity of about 220~mW/cm$^2$ is obtained when focused to a spot diameter of 3~$\mu$m, which is sufficient for driving nuclear Rabi oscillations. Although efforts are required to narrow down the mode bandwidth and improve the knowledge about the isomer's transition energy, no conceptual hindrance is seen (see Ref.~\cite{Wense2019} for details).\\[0.2cm]
Once narrow-band VUV lasers for the nuclear isomeric transition are available, one could also think of implementing a nuclear qubit for quantum computing. This was previously mentioned in  Ref.~\cite{Reader2011}, however without providing any further information. Here we briefly sketch future possibilities for this application. 
 Quantum optical control over the $^{229}$Th isomeric transition would bring with it the feasibility of a nuclear quantum bit.  Practically a Th-based quantum computer could be implemented as suggested in Ref. \cite{CiracZoller1995}, using Th ions in  linear Coulomb crystals. Such crystals with Th$^{3+}$ ions using the $^{229}$Th and  $^{232}$Th isotope species have been experimentally demonstrated in Refs. \cite{Campbell2011,Campbell2009}. The Th qubits would operate between the hyperfine sublevels $\left| g \right\rangle$ and $\left| e \right\rangle$ of the nuclear ground and isomeric states, respectively. The very low radiative decay rate and absence of other decay channels of the upper state in Th$^{3+}$ would prevent significant decoherence in such system. Apart from single qubit manipulations, a controlled-NOT quantum gate based on entanglement of two qubits is necessary for quantum computations \cite{SleatorWeinfurter1995}. This can be achieved via the centre-of-mass mode of the collective motion of the Coulomb crystal using an auxiliary excited state $\left| s \right\rangle$ \cite{CiracZoller1995} which in our case can be chosen from the hyperfine structure of the nuclear isomer.

\section{\label{sec9}The case of $^{235\text{m}}$U}

\noindent Another isotope with a low-energy nuclear excited state is $^{235}$U. With an energy of 76.73~eV \cite{Ponce}, its metastable state, $^{235\text{m}}$U, possesses the second lowest nuclear excitation energy known to date. This low excitation energy has led to the conclusion that also $^{235\text{m}}$U could be a candidate for direct nuclear laser excitation, as soon as laser technology has evolved to deliver intense laser light at the required wavelength. In this case, the long IC half-life of $^{235\text{m}}$U of about 26~minutes would even make the isomer a good candidate for a nuclear clock.\\[0.2cm]
Such considerations, however, do not take the multipolarity of the $^{235}$U isomer- to ground state transition into account. Being a transition from the spin and parity values $1/2^+$ (excited state) to $7/2^-$ (ground state), the multipolarity is $E3$. Similar like Eq.~(\ref{A}) the radiative transition rate for electric multipole radiation with $L=3$ is
\begin{equation}
\label{235Ulifetime}
\Gamma_\gamma=\frac{8}{33075}\frac{k^7}{\hbar\epsilon_0}B_{\downarrow}(E3).
\end{equation}
The latest available literature value for the reduced transition probability is $B_{\downarrow}(E3)=0.036$~W.u. \cite{Berengut}. One Weisskopf unit for an electric transition of arbitrary multipole order $L$ is defined as \cite{Weisskopf}
\begin{equation}
1\ \mathrm{W.u.}(EL)=\frac{1}{4\pi}\left(\frac{3}{3+L}\right)^2R^{2L}e^{2}.
\end{equation}
Here $e=1.602\cdot10^{-19}$~C is the electron charge. Inserting $L=3$ for $^{235\text{m}}$U, one obtains $1\ \mathrm{W.u.}=0.166$~fm$^6e^2$. Using this value together with $k=3.87\cdot10^8$~m$^{-1}$ to calculate the expected radiative decay rate based on Eq.~(\ref{235Ulifetime}) the result is $\Gamma_\gamma\approx9.4\cdot10^{-25}$~s$^{-1}$, corresponding to a lifetime of about $1\cdot10^{24}$~s ($3.4\cdot10^{16}$ years). The reason for the actually observed lifetime $\tau=t_{1/2}/\ln(2)$ of about 37.5~min \cite{Freedman} ($1.56\cdot10^3$~s) is a huge IC coefficient of $\alpha_{ic}\approx4.7\cdot10^{20}$. According to Eq.~(\ref{I1}) the laser intensity required to drive Rabi oscillations with a single tooth of an XUV frequency comb of bandwidth $\Gamma_\ell= 2\pi\cdot 1$~Hz can be estimated to be
\begin{equation}
I_\ell\gg\frac{\hbar\omega_0^3\Gamma_\ell^2}{8\pi c^2\Gamma_\gamma}\approx 3.0\cdot10^{20}\ \text{W}/\text{cm}^2.
\end{equation}
Here the angular frequency $\omega_0=1.15\cdot10^{17}$~Hz was used. Although focused laser intensities of beyond $10^{20}$ W/cm$^2$ have already been achieved for few-ten fs pulse durations at 800~nm, these laser systems possess enormously larger bandwidths. The calculated laser intensity of $3.0\cdot10^{20}$~W/cm$^2$ at a bandwidth of $2\pi\cdot 1$~Hz appears to be prohibitively large for the direct laser excitation of $^{235\text{m}}$U. As discussed in the previous section, the maximum reported XUV frequency comb intensity amounts to 220~mW/cm$^2$ at 150~nm, which is by about 21 orders of magnitude lower than the minimum intensity required to drive Rabi oscillations of $^{235\text{m}}$U. The problem gets even worse when considering that a higher harmonic order would be required for excitation. Nevertheless, laser excitation of $^{235\text{m}}$U could potentially be achieved via a sophisticated electronic bridge scheme in the $7+$ charge state \cite{Berengut}.

\section{\label{sec10}Summary}
\noindent A detailed discussion of the theory of direct nuclear laser excitation using the density matrix formalism is presented. Following an introductory part, in Sec.~\ref{sec2} the optical Bloch equations, Eq.~(\ref{twolevelbloch}), for a nuclear two-level system are derived. The explicit form of the Rabi frequency is obtained in Sec.~\ref{secRabi}, resulting in the expression given in Eq.~(\ref{Omega01}). Analytic solutions to the two-level optical Bloch equations, Eq.~(\ref{twolevelbloch}), are presented in Sec.~\ref{sec5} for three different cases: the case of zero detuning leads to Torrey's solution given in Eq.~(\ref{Torrey}), the low-saturation limit results in Eq.~(\ref{lowsat2}) and the steady-state case is solved by Eq.~(\ref{power}). Building up on these didactic parts, in Sec.~\ref{sec3} the complete set of differential equations describing a nuclear two-level system with HFS is presented. The section is split into two parts: the first part describes the nuclear HFS for free atoms and ions, resulting in the optical Bloch equations given in Eq.~(\ref{bloch60}), in the second part the splitting in a solid-state environment is discussed, leading to Eq.~(\ref{bloch51}). Both equations are then analytically solved for the low-saturation limit in Sec.~\ref{sec4}, resulting in Eq.~(\ref{Nt}). The theory of direct nuclear two-photon excitation is presented in Sec.~\ref{sec7}, leading to the result that nuclear excitation rates, described in Eq.~(\ref{twophotonrate}), can be generally considered as very low. The special cases of $^{229\text{m}}$Th and $^{235\text{m}}$U, being the nuclear excited states of lowest known excitation energies, are discussed in Sec.~\ref{sec8} and Sec.~\ref{sec9}. Direct nuclear laser excitation of $^{229\text{m}}$Th appears to be realistic with existing laser technology, while the required laser power to drive $^{235\text{m}}$U appears to be prohibitively large.

\section{Acknowledgement}
We acknowledge discussions with E. Peik, J. Thielking, T. Udem, J. Ye, C. Zhang, G. Porat, C.M. Heyl, S. Schoun and T. Schumm. L.v.d.W. wishes to thank the Humboldt foundation for financial support. This work was supported by DFG (Th956/3-2), by the Austrian Science Foundation (FWF) under grant No F41 (SFB ``VICOM''), and by the European Union's Horizon 2020 research and innovation programme under grant agreement 664732 ``nuClock".

\section{Appendix A: Averaging over fluctuations of the laser field}
\label{App:A}

\def\MM{\mathbb{M}}
\def\NN{\mathbb{N}}
\def\SS{\bar{S}}
\def\bro{\bar{\rho}}

\noindent The influence of phase fluctuations on the excitation of optical transitions has been studied in a number of works, e.g., in Refs.~\cite{Mazets92,Dalton82}. Here the respective analysis for the particular case of $\delta$-correlated phase fluctuations is presented, described by the equations
\begin{equation}
\langle \dot\phi(t) \rangle = 0; \quad 
\langle \dot\phi(t) \dot\phi(t') \rangle = \Gamma_\ell \delta(t-t').
\label{A:GammatotCoh}
\end{equation}
This $\delta$-correlated description takes account for the fact that the typical correlation times of phase fluctuations are significantly shorter than the typical evolution times of the system. The brackets denote the ensemble average and the correlation function defined as \cite{Riehle}
\begin{equation}
\langle b(t) \rangle= \text{lim}_{N\rightarrow\infty} \frac{1}{N}
\sum_{i=1}^{N} b_i(t)
\end{equation}
and
\begin{equation}
\langle b(t)b(t') \rangle= \text{lim}_{N\rightarrow\infty} \frac{1}{N}\sum_{i=1}^{N} b_i(t)b_i(t'),
\end{equation}
respectively. Different indices $i$ denote different systems in the ensemble. This correlation corresponds to a 
Lorentzian shape of the laser spectrum with ${\rm FWHM}=\Gamma_\ell$, as will be shown in the following.
First, an averaging of Eqs.~(\ref{bloch2}) over these fluctuations is performed. This set of equations is degenerate, but the degeneracy can be waived with the help of the normalization condition $\sum_i \rho_{ii}=1$. It can be written in matrix form as
\begin{equation}
\dot{\bro} = \left[ \MM + \NN (t) \right] \cdot \bro + \SS,
\label{A:MatrForm}
\end{equation}
where the column vector $\bro$ consists of matrix elements of the density matrix $\hat{\rho}$, the matrix $\MM$ is time-independent, $\SS$ is a constant vector appearing due to the normalization condition, and $\NN(t)$ is a fast fluctuating diagonal matrix proportional to $\dot{\phi}$, whose only non-zero elements are $i\dot{\phi}$ (corresponding to elements $\rho_{eg}$), and $- i\dot{\phi}$ (corresponding to $\rho_{ge}$):
\begin{equation}
\NN(t)= \dot{\phi}(t) \, \NN^0; \quad \NN^0_{eg,eg}=i; \quad \NN^0_{ge,ge}=-i\, .
\label{A:NNdef}
\end{equation}

\noindent The solution of Eq.~(\ref{A:MatrForm}) can be represented as a sum of two terms, slowly and rapidly varying in time, as in \cite{Romanenko}:
\begin{equation}
\bro(t) = \bro_0(t)+\bro^\prime(t).
\label{A:sol0}
\end{equation}
The slow term $\bro_0=\langle \bro_0 \rangle$ varies with a characteristic time of order of the inverse eigenvalues of the matrix $\MM$, and the fast term $\bro'$ varies with a characteristic time of order of the correlation time of the noise terms. It is adopted that the average value of the rapidly oscillating terms vanishes over times much longer than the correlation time of the phase noise terms, but much shorter than the characteristic evolution time of the slow terms; the same is correct for the ensemble average, i.e., $\langle \bro' \rangle=0$, and $\bro_0=\langle \bro_0 \rangle$. Then Eq.~(\ref{A:MatrForm}) gives
\begin{equation}
\dot{\bro}_0+\underline{\dot{\bro}'} = \MM \cdot \left[\bro_0+ \underline{\bro'}\, \right]
           + \underline{\NN (t) \left[\bro_0+ \underline{\bro'}\, \right]}  + \SS.
\label{A:MatrFormSloFast}
\end{equation}
The underlined terms fluctuate rapidly, and vanish after having been averaged over the ensemble, as well as over the time longer than the correlation time of the phase noise. The twice underlined term consists of a product of rapidly fluctuating terms and may contain a slowly varying component. This allows to write
\begin{align}
\dot{\bro}_0(t) = & \MM \cdot \bro_0(t) + \SS + \langle \NN(t) \bro'(t) \rangle, 
\label{A:eqSlow} \\
\dot{\bro}'(t)  = & \MM \cdot \bro'(t) + \NN(t) \bro_0(t).
\label{A:eqFast}
\end{align}
The solution of Eq. (\ref{A:eqFast}) can be written as
\begin{equation}
\begin{split}
\bro'(t)& = \int\limits_{-\infty}^t e^{\MM (t-t')} \cdot \NN (t') \cdot \bro_0(t') dt' \\
& \approx \int\limits_{-\infty}^t e^{\MM (t-t')} \cdot \NN (t')  dt' \cdot \bro_0(t),
\label{A:solFast}
\end{split}
\end{equation}
where it was used that $\bro_0$ is a slow variable. Substituting this into Eq.~(\ref{A:eqSlow}) and using Eqs. (\ref{A:GammatotCoh}) and (\ref{A:NNdef}), one obtains
\begin{equation}
\begin{aligned}
\dot{\bro}_0(t) &= \left[ \MM + 
\int\limits_{-\infty}^t  
    \left\langle \NN(t) \cdot e^{\MM (t-t')} \cdot \NN(t') \right\rangle dt'
\right]\cdot \bro_0(t) + \SS 
\\
&= \left[ \MM + 
\int\limits_{-\infty}^t  \left\langle \NN(t) \cdot \NN(t') \right\rangle dt'
\right]\cdot \bro_0(t) + \SS 
\\
&= \left[ \MM + 
\frac{\Gamma_\ell}{2}{\NN^0}^2 
\right]\cdot \bro_0(t) + \SS,
\\
\end{aligned}
\label{A:eqSlowMod1}
\end{equation}
or
\begin{equation}
\langle\dot{\bro}\rangle = \left[ \MM + \frac{\Gamma_\ell}{2}{\NN^0}^2 \right] 
 \cdot \langle{\bro}\rangle + \SS.
\label{A:MatrForm2}
\end{equation}

\noindent It follows from the form of the matrix $\NN^0$ [Eq.~(\ref{A:NNdef})] that averaging over the fluctuating laser phase leads to the appearance of additional relaxation terms $\Gamma_\ell/2$ for the non-diagonal matrix elements $\rho_{eg}$ and $\rho_{ge}$. 

\noindent To clarify the physical meaning of $\Gamma_\ell$, the power spectral
density $S_E(f)$ of the field is calculated:
\begin{equation}
\begin{aligned}
E(t)&=E_{0}\left(e^{-i(\omega_\ell t+\phi(t))}+
            e^{i(\omega_\ell t+\phi(t))}\right)\\
&=2E_{0} \cos(\omega_\ell t+\phi(t)).
\end{aligned}
\label{A:Ext}  
\end{equation}
According to the Wiener-Khinchin theorem \cite{Riehle}, the power spectral density $S_E(f)$ of the signal $E(t)$ may be expressed through its autocorrelation function $R_E(\tau)$ as
\begin{align}
S_E(f)=&\int\limits_{-\infty}^{\infty} e^{-2 \pi i f \tau} R_{E}(\tau) d \tau, \quad 
      {\rm where}\\
R_E(\tau)=&\langle E(t) E(t+\tau) \rangle.
\label{A:Spectr}  
\end{align}
Here $f=2\pi \omega$ is the ordinary frequency. Suppose that the phase fluctuations are Gaussian, then, using the relation for Gaussian random phases
\begin{equation}
\langle \exp\left[i(\phi(t)-\phi(t^\prime))\right] \rangle = \exp\left[-\frac{\langle(\phi(t)-\phi(t^\prime))^2\rangle}{2}\right],
\end{equation}
after some algebra, the autocorrelation function $R_E(\tau)$ can be expressed as
\begin{equation}
R_E(\tau)= 2 E_{0}^2 \exp\left[-\frac{\Gamma_\ell |\tau|}{2}\right] \cos(\omega_\ell \tau).
\label{A:Autocorr}  
\end{equation}
Substituting Eq.~(\ref{A:Autocorr}) into Eq.~(\ref{A:Spectr}), the power spectral density $S_E(f)$ is obtained in the form:\\
\begin{multline}
S_E(f)=2 E_0^2
    \left[ 
        \frac{\Gamma_\ell/2}{
           \left(
                \frac{\Gamma_\ell}{2}\right)^2+\left( 2 \pi f + \omega_\ell 
           \right)^2} \right. \\
+ \left. 
        \frac{\Gamma_\ell/2}{
           \left(
                \frac{\Gamma_\ell}{2}\right)^2+\left( 2 \pi f - \omega_\ell 
           \right)^2}        
    \right].
\end{multline}
Therefore, the phase fluctuations given in Eq.~(\ref{A:GammatotCoh}) correspond to a Lorentzian spectrum, and $\Gamma_\ell$ is the full linewidth at half maximum.


\section{Appendix B: Derivation of the optical Bloch equations for a solid-state environment}
As usual, for the derivation of the optical Bloch equations, the quantum optical master equation in Lindblad form, Eq.~(\ref{master}), is used as the starting point. For the considered case of the solid-state environment, the Lindblad superoperator $\mathcal{L}[\hat{\rho}]$ has a more complicated form, since it describes, besides the spontaneous decay of the excited level sub-states to the ground level sub-states, also various other relaxation processes due to interaction with the environment. It can be represented as
\begin{equation}
\LL[\hat{\rho}]=\LL_{\rm sp}[\hat{\rho}]+\LL_{\rm od}[\hat{\rho}]+\mathcal{L}_\text{mix,e}[\hat{\rho}]
                       +\mathcal{L}_\text{mix,g}[\hat{\rho}]+\LL_{\rm md}[\hat{\rho}],
\label{lindbladsuper}
\end{equation}
where $\LL_{\rm sp}$ contains spontaneous decay of the excited states to the ground states and was already defined in Eq.~(\ref{lindbladsp20}). $\LL_{\rm od}$ models the loss of coherence between excited and ground states due to some random process shifting the levels with respect to each other. In the considered example this is introduced by the laser light with bandwidth $\Gamma_\ell$ and $\LL_{\rm od}$ takes the form
\begin{multline}
\LL_{\rm od}[\hat{\rho}]= - \frac{\Gamma_\ell}{4}  
\left[ \hro  - 
\vphantom{\left(\sum_e \sig_{ee} - \sum_g \sig_{gg} \right)}
\right.
\\
\left. \left(\sum_e \sig_{ee} - \sum_g \sig_{gg} \right) \hro 
\left(\sum_e \sig_{ee} - \sum_g \sig_{gg} \right) \right].
\label{lindbladod}
\end{multline}
$\mathcal{L}_\text{mix,e}[\hat{\rho}]$ and $\mathcal{L}_\text{mix,g}[\hat{\rho}]$, respectively, describe transitions between different magnetic sub-states due to non-controllable fluctuations of the local magnetic field created, e.g., by spin changes in the environment as well as variations of the local electric field gradient caused by thermal motion of the atoms in the solid-state environment. The population mixing between different sublevels of the ground ($i,i'=g,g'$) or excited ($i,i'=e,e'$) states is defined by the Lindblad operator
\begin{equation}
\mathcal{L}_\text{mix, i}[\hat{\rho}]=\sum_{i, i'}\gamma_{ii'}\left(\hat{\sigma}_{ii'}\hat{\rho}\hat{\sigma}^\dagger_{ii'}-\frac{1}{2}\hat{\rho}\hat{\sigma}^\dagger_{ii'}\hat{\sigma}_{ii'}-\frac{1}{2}\hat{\sigma}^\dagger_{ii'}\hat{\sigma}_{ii'}\hat{\rho}\right),
\label{Lmixi}
\end{equation}
where $\gamma_{gg'}$ and $\gamma_{ee'}$ denote the decay rates between different sub-states of the same level. $\LL_{\rm md}$ takes account for loss of coherence between different magnetic sub-states caused by the same reasons. This can be described by
\begin{equation}
\begin{split}
\LL_{\rm md}[\hat{\rho}]=&\sum_{g\neq g^\prime}\frac{\tilde{\gamma}_{gg'}}{2}
   \left[ 
       (\sig_{gg}-\sig_{g'g'})\hro(\sig_{gg}-\sig_{g'g'}) 
        \vphantom{\frac{1}{2}}
   \right.
       \\
& \hphantom{aaaaaaaaa} \left. - \frac{1}{2}
           \left\{ (\sig_{gg}+\sig_{g'g'}),\hro  \right\}
   \right] \\
&+ \sum_{e\neq e^\prime}\frac{\tilde{\gamma}_{ee'}}{2}
   \left[ 
       (\sig_{ee}-\sig_{e'e'})\hro(\sig_{ee}-\sig_{e'e'}) 
        \vphantom{\frac{1}{2}}
   \right.
       \\
& \hphantom{aaaaaaaaa} \left. - \frac{1}{2}
           \left\{ (\sig_{ee}+\sig_{e'e'}),\hro  \right\}
   \right],   
\label{lindbladmd}
\end{split}
\end{equation}
where $\tilde{\gamma}_{gg'}$ and $\tilde{\gamma}_{ee'}$ denote the decoherence rates between different sub-states. The complete set of sub-state optical\newpage \noindent Bloch equations in case of laser excitation in a solid-state environment is then obtained in the form \cite{Kazakov1,Kazakov2}\\

\begin{strip}
\vspace{1.5cm}
\begin{equation}
\begin{aligned}
\dot{\rho}_{ee}=& -i \sum_{g} \frac{\Omega_{eg}}{2}
        \Big[ \rho_{ge}-\rho_{eg}\Big] -
        \sum_{g} \Gamma_{ge} \rho_{ee} 
        + \sum_{e'} \gamma_{ee'} (\rho_{e'e'}-\rho_{ee}); 
\\
\dot{\rho}_{ge}=&-i\sum_{{e'}}  \frac{\Omega_{e'g}}{2} \rho_{e'e}
          +i\sum_{{g'}} \frac{\Omega_{eg'}}{2} \rho_{gg'}
          -\rho_{ge}(i\Delta\omega_{eg}+\tilde{\Gamma}_{ge});
\\
\dot{\rho}_{gg}=& -i \sum_{e} 
        \frac{\Omega_{eg}}{2}\Big[\rho_{eg}-\rho_{ge}\Big] +
        \sum_{e} \Gamma_{ge} \rho_{ee}
         + \sum_{g'} \gamma_{gg'} (\rho_{g'g'}-\rho_{gg}); 
\\
\dot{\rho}_{ee'}=&  -i \sum_{g}
        \Big[ \frac{\Omega_{eg}}{2} \rho_{ge'}- \frac{\Omega_{e'g}}{2} \rho_{eg} \Big] 
        - \rho_{ee'} (i \omega_{ee'} + \tilde{\Gamma}_{ee'}), \quad e \neq e';
\\
\dot{\rho}_{gg'}=& -i \sum_{e}
        \Big[ \frac{\Omega_{eg}}{2} \rho_{eg'}- \frac{\Omega_{eg'}}{2} \rho_{ge} \Big]
        -\rho_{gg'}(i \omega_{gg'} + \tilde{\Gamma}_{gg'}), \quad g \neq g'.
\label{bloch5}
\end{aligned}
\end{equation}
\end{strip}

\noindent As previously, $\omega_{ij}$ and $\omega_{eg}$ were defined as $\omega_{ij}=\omega_i-\omega_j$ and $\Delta\omega_{eg}=\omega_\ell-\omega_{eg}$. This time, the decay rates of the coherences are obtained in the form:
\begin{equation}
\label{gammatilde1}
\begin{aligned}
\tilde{\Gamma}_{ge}&=\sum_{g'}\frac{\Gamma_{g'e}+\gamma_{g'g}}{2}
                    +\sum_{e'}\frac{\gamma_{e'e}}{2}+\frac{\Gamma_\ell}{2}  \\
                    & +\sum_{g''\neq g}\frac{\tilde\gamma_{g g''}+\tilde\gamma_{g'' g}}{4}
                    +\sum_{e''\neq e}\frac{\tilde\gamma_{e e''}+\tilde\gamma_{e'' e}}{4},\\
\tilde{\Gamma}_{ee'}&=\sum_{g} \frac{\Gamma_{ge}+\Gamma_{ge'}}{2}
                      +\sum_{e''}\frac{\gamma_{e''e}+\gamma_{e''e'}}{2} \\
                      & +\frac{\tilde\gamma_{ee'}+\tilde\gamma_{e'e}}{2}+
                      \sum_{e''\neq e} \tilde \gamma_{ee''}
                      +\sum_{e''\neq e'} \tilde \gamma_{e''e'},\\
\tilde{\Gamma}_{gg'}&=\sum_{g''}\frac{\gamma_{g''g}+\gamma_{g''g'}}{2}
                      +\frac{\tilde\gamma_{gg'}+\tilde\gamma_{g'g}}{2}\\
                      &+\sum_{g''\neq g} \tilde \gamma_{gg''}
                      +\sum_{g''\neq g'} \tilde \gamma_{g''g'}.
\end{aligned}
\end{equation}
In the case that $\Omega_{eg}^2/\tilde{\Gamma}_{ge}$ is small compared to the energy differences $\omega_{gg'}$ and $\omega_{ee'}$, or in comparison with the decoherence rates $\tilde{\Gamma}_{ee'}$ and $\tilde{\Gamma}_{gg'}$, again it is possible to neglect the two-photon coherences $\rho_{ee'}$ and $\rho_{gg'}$ and Eq.~(\ref{bloch5}) transforms to Eq.~(\ref{bloch51}).


\end{document}